\documentclass[aip,reprint,twocolumn]{revtex4-1}
\usepackage{amsmath}
\usepackage{graphicx}
\usepackage{natbib}
\usepackage{subfigure}
\usepackage{color}
  
\begin{document}

\title{Evaporation of dilute droplets in a turbulent jet: clustering and entrainment effects}

\author{Federico Dalla Barba}
\author{Francesco Picano}
\affiliation{University of Padova, Department of Industrial Engineering, Via Venezia 1, 35131 Padova, Italy}

\date{\today}

\begin{abstract}

Droplet evaporation in turbulent sprays involves unsteady, multiscale and multiphase processes which make its comprehension and model capabilities still limited. The present work aims to investigate droplet vaporization dynamics within a turbulent spatial developing jet in dilute, non-reacting conditions. We address the problem using a Direct Numerical Simulation of jet laden with acetone droplets using an hybrid Eulerian/Lagrangian approach based on the point droplet approximation. A detailed statistical analysis of both phases is presented. In particular, 
we show how crucial is the preferential sampling of the vapour phase induced by the inhomogeneous localization of the droplets through the flow. 
The preferential segregation of droplets develops suddenly downstream the inlet both within the turbulent core and in the mixing layer.
Two distinct mechanisms have been found to drive these phenomena, the inertial small-scale clustering in the jet core and the intermittent dynamics of droplets across the turbulent/non-turbulent interface in the mixing layer where dry air entrainment occurs. 
These phenomenologies strongly affect the overall vaporization process and lead to a spectacular widening of droplets size and vaporization rate distributions
 in the downstream evolution of the turbulent spray. 
\end{abstract}

\maketitle  

\section{Introduction}

Turbulent sprays are complex multiphase flows involving unsteady and multi-scale phenomena such as turbulence coupled with phase transition processes. The presence of two distinguished phases which mutually interact exchanging mass, momentum and energy makes the description of the problem extremely challenging. If combustion is considered, chemical reactions and heat release add some clear complexities. In this scenario, a satisfactory comprehension of turbulent spray dynamics has not yet been achieved and existing models capabilities for applications are still limited~\cite{jenny2012,kruger2016effect}. Nevertheless, the progress of the research in this field is crucial for several industrial applications as well by an environmental point of view. A typical example can be found in the development of high efficiency and low emission internal combustion engines. In these applications the liquid fuel can be directly injected  into the combustion chamber where the vaporization of fuel droplets occurs together with chemical reactions within the turbulent gaseous environment.  The pollutants formation in turbulent spray combustion is related to complex multi-scale phenomenons that involve fluctuations of temperature and reactants concentrations. In particular, the soot formation occurs through a pyrolysis process in fuel-rich regions that experience high temperature without enough oxidizer to react ~\cite{raman2016modeling,attili2014formation,kennedy1997models}. This can be observed within droplets clusters, where the concentration of fuel droplets can be even thousand times higher then its bulk value leading to a peak in the fuel vapor concentration. Hence, in order to predict and model soot formation an improved understanding of the mechanisms that govern the distribution of droplets and fuel/oxidizer mixture within a turbulent jet spray is mandatory.

A phenomenological description of the overall evolution of the spray dynamics can be found in the review of Jenny et al. ~\cite{jenny2012}. The process starts with the \emph{primary atomization} of a high velocity liquid jet. As the liquid flow is ejected from a duct into a gaseous environment, interface instabilities such as Kelvin-Helmholtz and Rayleigh-Taylor, fragmentize the jet into large drops and ligaments ~\cite{marmottant2004spray}. In the downstream evolution the liquid ligaments and the drops are subjected to aerodynamic forces induced by the velocity difference at liquid/gas interface. The stresses induced by the aerodynamic forces produce a further brake-up of the liquid phase (\emph {secondary atomization}) giving origin to a system of small droplets dispersed within the gaseous turbulent phase. The atomization process occurs in a so called \emph{dense regime} and terminate when surface tension prevails on aerodynamic stresses preventing further fragmentation. At this step a \emph{dilute regime} establishes: droplets mutual interactions (e.g. collisions and coalescence) are negligible but the effect of droplets on the carrier flow is still significant~\cite{ferrante2003physical,gualtieri2015exact}. While in the dense regime the vaporization rate is negligible, in dilute conditions the vaporization process becomes significant. In this phase of the turbulent spray evolution, the most part of the liquid evaporates. Moreover, the small droplets evolve preserving a spherical shape due to the dominance of surface tension on aerodynamic stresses. Even if the presence of droplets exerts a significant effect on the flow in terms of mass, momentum and energy balance, at this step the dimension of droplets is below or comparable to the smallest scales of the turbulent flow so point-droplet approximation applies~\cite{elghobashi1994predicting}. Hence, in dilute conditions the mathematical description of droplet-laden flows lends itself particularly well to an hybrid Eulerian/Lagrangian description. The Navier-Stokes equations govern the continuous phase dynamics if distributed sink-source terms are considered in order to represent the mass, momentum and energy exchange between the Eulerian carrier phase and Lagrangian point-droplets. 

One of the most peculiar phenomenology that characterize dispersed multiphase turbulent flows in dilute conditions is the preferential segregation of particles/droplets as a result of the interaction of their inertia with the carrier phase turbulent dynamics (see e.g. ~\cite{toschi2009lagrangian}). The mechanisms at the origin of preferential segregation in free flows have been widely investigated in literature both for solid particles and liquid droplets. The intensity of small scale segregation of solid particles in homogeneous turbulence is found to be driven by the Stokes number $St_{\eta}=\tau_p/\tau_{\eta}$, with $\tau_p$ the particle relaxation time and $\tau_{\eta}$ the Kolmogorov time scale. More specifically, the intensity of small scale clustering is maximum when the particle relaxation time is of the order of the Kolmogorov time scale such that the Stokes number, $St_\eta\simeq1$ \cite{calzavarini2008quantifying, bec2007heavy, saw2008inertial, gualtieri2009anisotropic}. The same behavior is observed for evaporating droplets which behave as inertial particles\cite{reveillon2007effects}. Concerning turbulent jets, a mean accumulation of the dispersed phase has been observed at specific distances from the inflow both experimentally ~\cite{lau2016effect} and numerically ~\cite{picano2010anomalous}. Even if  preferential concentration of a dispersed phase has been well characterized in homogeneous turbulence, the effect of this phenomenology on the overall vaporization process within turbulent jet sprays is still object of research and constitute one of the main focuses of this paper.  

By a theoretical and numerical point of view one of the first description of the vaporization of spherical droplets dragged by a gaseous phase flow was addressed in the seminal works of Spalding and Godsave ~\cite{spalding1953, godsave1953}. Fixing the environmental vapor concentration, they found that droplets surface decreases linearly with time ($D^2$ law). Maxey \& Riley, ~\cite{maxey1983equation} report an equation for the motion of a finite size spherical particle/droplets at low droplet Reynolds number. The equation accounts for the Stokes drag,  added-mass effect and buoyancy force. Dealing with a dispersed phase with a density much higher than that of the fluid the Stokes drag is sufficient to describe its dynamics ~\cite{armenio2001importance, olivieri2014effect}.  Abramzon \& Sirignano ~\cite{abramzon1989droplet} proposed an improved model for droplets vaporization, considering non-uniform and time-dependent environmental conditions, taking into account forced convection, molecular diffusion and the Stefan flow contribution to the vapor transport from droplet surface to neighboring environment. Even if several models ~\cite{jenny2012, marchioli2008some} can be found in literature for the simulation of turbulent evaporating sprays in Reynolds Averaged Navier-Stokes (RANS) or Large-Eddy Simulation (LES) frameworks, these models lack in capabilities to accurately reproduce complex phenomena such as droplets small scale clustering~\cite{kruger2016effect}. Despite the highly demanding computational resources, the use of Direct Numerical Simulation (DNS) allows to capture the whole physics of the spray vaporization process in order to understand the complex phenomenons involved. In this context,  Mashayek ~\cite{mashayek1998direct} adopted an Eulerian/Lagrangian approach in order to perform a DNS of low Mach number, homogeneous shear turbulent flows laden with droplets . Miller \& Bellan ~\cite{miller1999direct} report a DNS of a confined three-dimensional, temporally developing gas mixing layer laden with evaporating hydrocarbon droplets at subsonic Mach number. Reveillon and co-workers ~\cite{reveillon2007effects} studied the effect of preferential droplet accumulation on the evaporation in isotropic turbulence showing that different regime takes place depending on droplet concentration. 
Recently, Bukhvostova et al. ~\cite{bukhvostova2014comparison} consider the DNS of a turbulent channel flow of a mixture of air and water vapor laden with water droplets. The work focuses on the comparison between the performances of an incompressible and a low Mach number asymptotic formulation in reproducing the flow dynamics. Even if the two formulations show good qualitative agreement, the low Mach formulation is found to be crucial in order to obtain a reliable quantitative predictions of heat and mass transfer.

The prototypal flow for an evaporating spray is constituted by a turbulent free jet which is characterized by the effect of environmental gas entrainment. In more details, a turbulent jet is constituted by a rotational turbulent core which is continuously entrained by the surrounding irrotational fluid ~\cite{da2014interfacial}. In sprays, the entrained dry flow dilutes the vapor concentration and controls the vaporization process. This phenomenology was found to be of critical importance also in natural phenomena. One example consists in the effect of entrainment in stratocumulus clouds where it constitutes a driving parameters in the determination of cloud lifetimes and  in turns even regulating planetary-scale properties ~\cite{mellado2017cloud}. The fast grow rate of the droplet size spectrum in warm cloud is a challenging, still not understood, problem in cloud physics ~\cite{falkovich2002acceleration, sardina2015continuous}, despite its importance in determining raining conditions. 

To the best of authors knowledge, a fundamental study on the effects of the entrainment in an evaporating turbulent jet spray together with an analysis of  the preferential segregation effect is still missing. This work aims to cover this lack considering DNS data of an evaporating turbulent spray using a 2-way coupling approach between the two phases and accounting for the entrainment effect. The  numerical algorithm adopts a hybrid Eulerian/Lagrangian approach and point droplets approximation. In addition, the effect of density variation is accounted by a low Mach number formulation of the Navier-Stokes equations.  A strong preferential segregation of droplets is observed over the whole downstream evolution of the spray which induces a preferential sampling of vapor concentrated regions. Two different mechanisms are found to drive this process, the former is due to  inertial clustering, the second is related to the dynamics of the jet 
entrainment. This last mechanism is found to be crucial in the outer part of the jet core where the evaporation peaks and strongly impacts the vaporization dynamics which is characterized by a strong widening of the droplet size spectrum. 

\section{Numerical Method}
In this paper we report a direct numerical simulation of a turbulent evaporating spray in an hybrid Eulerian/Lagrangian framework adopting the point droplet approximation in the 2-way coupling conditions. The governing equations for the Eulerian gaseous phase consist in a low Mach number formulation of the Navier-Stokes equations in an open environment where arbitrary density variations can be accounted neglecting acoustics~\cite{majda1985derivation}. Consistently with previous studies ~\cite{miller1999direct, bukhvostova2014comparison}, the effect of the dispersed phase on the gaseous phase is accounted by sink/source coupling terms appearing in the right hand side of the mass, momentum and energy equations, $S_\rho$, $\bf{S_m}$ and $S_e$ (see ~\cite{mashayek1998direct}),

\begin{align}
\label{eq:1}
& \frac{\partial \rho}{\partial t}+\nabla \cdot (\rho \mathbf{u})=S_\rho
\\
& \frac{\partial}{\partial t}(\rho Y_V)+\nabla \cdot (\rho Y_V \mathbf{u}) = \nabla \cdot (\rho {\cal D} \nabla Y_V)+S_\rho
\\
& \frac{\partial}{\partial t}(\rho \mathbf{u})+\nabla \cdot (\rho \mathbf{u} \otimes \mathbf{u}) = \nabla \cdot \pmb{\tau} - \nabla {P} + \mathbf{S_m} 
\\
& \nabla \cdot \mathbf{u} = \frac{1}{p_0} \left[ \frac{\gamma-1}{\gamma} \nabla \cdot (k \nabla T) + \ S_e \right]
\\
\label{eq:2}
& p= \rho \  R_G (1+M \ Y_V) T
\end{align}

where $\rho$, $\bf u$, $p_0$, $P$ and $T$ are respectively the density, the velocity, the thermodynamic pressure, the hydrodynamic pressure and the temperature of the carrier vapor/gas mixture that will be denoted in the following as the carrier phase. The vapor mass fraction, $Y_V$, is defined as the ratio of the vapor partial density and the total density of the carrier phase, $Y_V=\rho_V/\rho$. The vapor/gas binary diffusion coefficient, the thermal conductivity of the mixture and the specific constant of gas are denoted respectively as $\cal D$, $k$ and $R_G$. The parameter $M$ is defined as $M=W_G/W_L-1$ with $W_G$ and $W_L$ the molar weight of the gas and liquid phases. The ratio of the constant pressure coefficient, $C_P$, and the constant volume coefficient, $C_V$, of the mixture is denoted by $\gamma$. The viscous stress tensor is $\pmb{\tau} = 2 \mu (\nabla \mathbf{u} + {\nabla \mathbf{u}}^T)-2\mu_b/3 \nabla \cdot \mathbf{u} \mathbf{I}$, with $\mu$ and $\mu_b$ the dynamic and bulk viscosities.  It should be remarked that the thermodynamic pressure $p_0$ is constant in space, due to the low-Mach number asymptotic expansion~\cite{majda1985derivation} and in time, due to the open space conditions.

Droplets are treated as rigid evaporating spheres and the liquid phase properties (e.g. temperature) are assumed to be uniform inside the droplets. Droplet rotation, distortion and mutual interactions (e.g. collision, coalescence) are neglected considering dilute volume fractions. 

\begin{align}
\label{eq:3}
& \frac{d \mathbf{u_d}}{dt}=\frac{\mathbf{u}-\mathbf{u_d}}{\tau_d}
\\
& \frac{dm_d}{dt}=-\frac{1}{3}\frac{m_d}{\tau_d} \frac{Sh}{Sc} \ ln(1+B_m)   
\\
\label{eq:4}
& \frac{dT_d}{dt} = \frac{1}{3 \ \tau_d} \left(\frac{Nu}{Pr} \frac{C_{P,G}}{C_L}(T-T_d)- \frac{Sh}{Sc} \frac{L_V}{C_L} \ ln(1+B_m)\right)
\end{align}

where $\mathbf{u_d}$, $m_d$ and $T_d$ are droplet velocity, mass and temperature, $C_{P,G}$ and $C_L$ are the constant pressure coefficient of the gas and the liquid specific heat and $\tau_d=2 \rho_L r_d^2 / (9 \mu)$ is the droplet relaxation time, with $\rho_L$ the liquid phase density. The mass diffusivity and the thermal conductivity are accounted through the Schmidt and Prandtl numbers respectively, $Sc=\mu / (\rho \cal D)$ and $Pr=\mu / (C_p k)$. The Nusselt number, $Nu_0$, and the Sherwood number, $Sh_0$, are estimated as a function of the droplets Reynolds number, $Re_d=\rho || \pmb{u} - \pmb{u}_d || / \mu $, according to the Fr\"{o}ssling correlation:

\begin{align}
& Nu_0 = 2+0.552 \ Re_d^{\frac{1}{2}} \ Pr^{\frac{1}{3}} 
\\
& Sh_0 = 2+0.552 \ Re_d^{\frac{1}{2}} \ Pr^{\frac{1}{3}}
\end{align}

A correction is then applied to $Nu_0$ and $Sh_0$ in order to account for the Stefan flow \cite{abramzon1989droplet}:

\begin{align}
& Nu = 2+\frac{(Nu_0-2)}{F_T},  & F_T= \frac{(1+B_t)^{0.7}}{B_t} \ ln(1+B_t)
\\
& Sh = 2+\frac{(Sh_0-2)}{F_M},  & F_M= \frac{(1+B_m)^{0.7}}{B_m} \ ln(1+B_m)
\end{align}

The parameter $B_m$ and $B_t$ are the Spalding mass and heat transfer number respectively, the former being the driven parameter for the vaporization rate,

\begin{align}
& B_m=\frac{(Y_{V,s}-Y_V)}{(1-Y_{V,s})}
\\
& B_t=\frac{C_{P,V}}{L_V}(T-T_d)
\end{align}

where $C_{P,V}$ is the constant pressure coefficient of the vapor, $Y_V$ is the vapor mass fraction in the carrier phase evaluated at droplet center and $Y_{V,s}$ is the vapor mass fraction evaluated at droplet surface. This latter corresponds to mass fraction of vapor  in a saturated vapor/gas mixture at droplet temperature. In order to estimate $Y_{V,s}$ we assume the equilibrium hypothesis such that the Clausius-Clapeyron relation applies:
 
\begin{equation}
\chi_{V,s}=\frac{p_{ref}}{p_0} exp \left[ { \frac{L_V}{R_V} \big( \frac{1}{T_{ref}}-\frac{1}{T_d} \big)} \right]
\end{equation}

with $\chi_{V,s}$ the vapor molar fraction at droplet surface, p the thermodynamic pressure, $p_{ref}$ and $T_{ref}$ arbitrary reference pressure and temperature and $R_V$ the vapor gas constant. The saturated vapor mass fraction is then:

\begin{equation}
Y_{V,s}=\frac{\chi_{V,s}}{\chi_{V,s}+(1-\chi_{V,s}) \frac{W_g}{W_L}}
\end{equation} 



\begin{figure}[h]
\subfigure[]{\includegraphics[width=0.47\columnwidth]{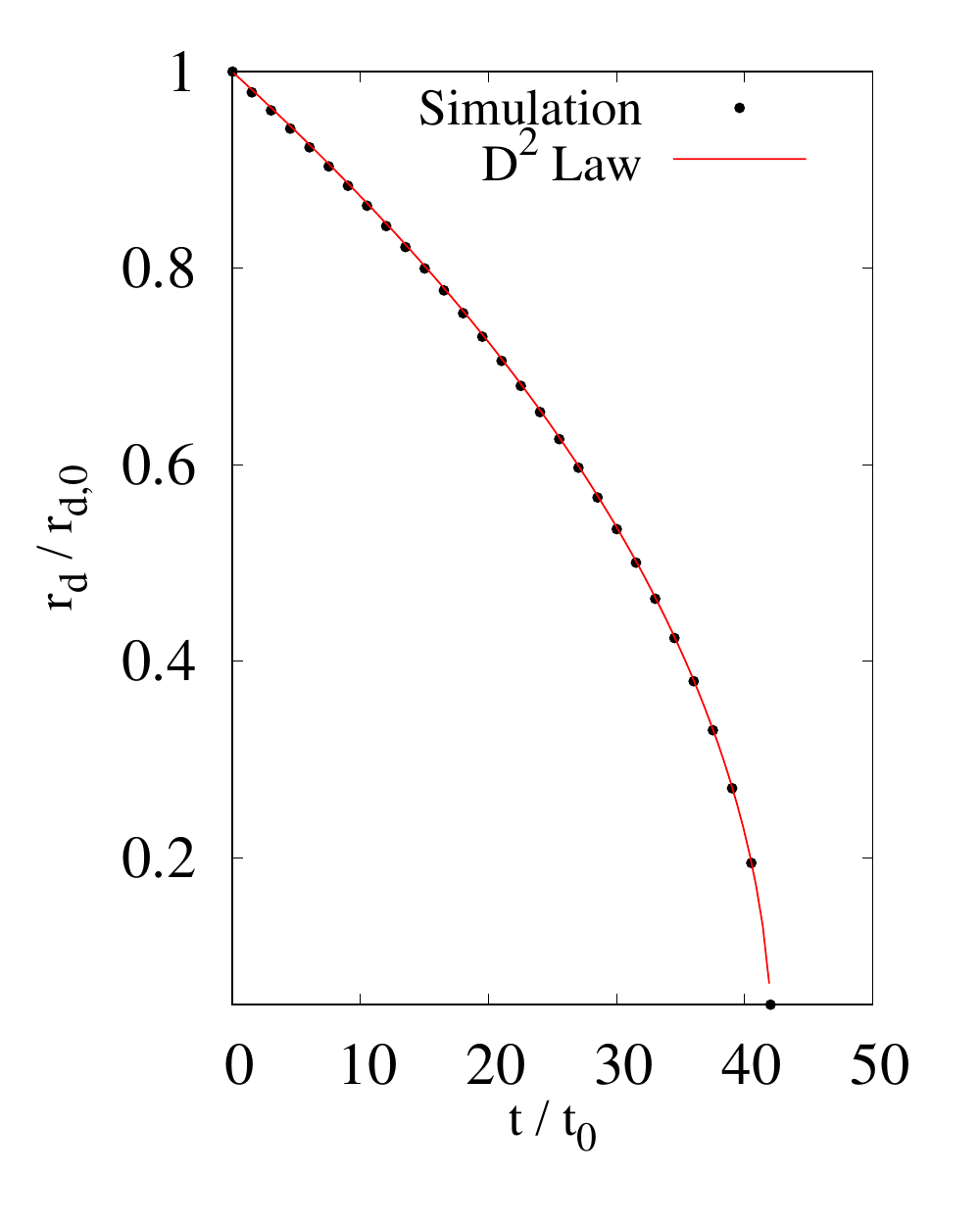} \label{fig:1:a}} \
\subfigure[]{\includegraphics[width=0.47\columnwidth]{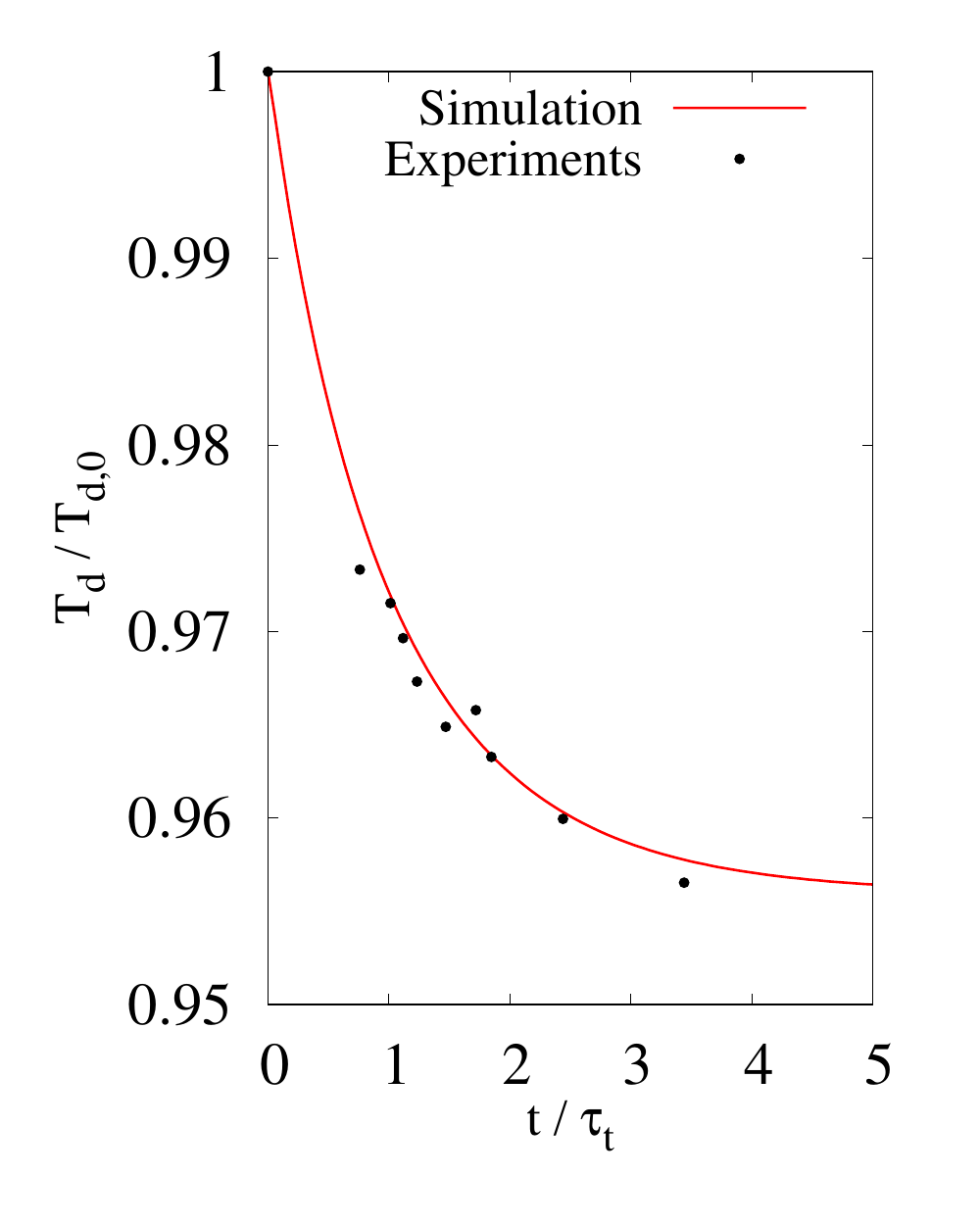} \label{fig:1:b}} \
 \caption{(a) Evolution over time of the radius of a single water droplet in a dry air laminar jet. The ambient pressure and temperature are set to $p= 101300 Pa$ and $T=273.15 K$. The inlet radius and the bulk velocity of the jet are respectively $R=5 \ 10^{-3} m$ and $U=1.7 m/s$. The droplet is injected at the local carrier phase velocity. The initial droplet radius is set to $r_{d,0}=5 \mu m$ and the temperature to $T_{d,0}=273.15 K$.  In the figure, the droplet radius and the time are scaled respectively by the droplet initial radius and the reference time scale $t_0=R/U$. The continuous line represents the analytical solution computed by the Spalding $D^2$ law, $r_d=\sqrt{r_{d,0}^2-k t}$, where $k=2 \rho {\cal D} \ ln(1+B_m) / \rho_L$. Dots represents the numerical results. (b) Temperature evolution over time of a water droplet freely falling in air at pressure $p=101300 Pa$, temperature $T=301.45 K$ and relative humidity $\chi=0.22$. Droplet is initially at rest with an initial temperature, $T_{d,0}$, equal to the environmental air one. In the figure the temperature is scaled by the initial droplet temperature, $T_{d,0}$, while time is scaled by droplet thermal relaxation time, $\tau_t$. This latter is defined as the time required by droplet temperature to change by the 63 \% of its total change between initial temperature and regime temperature. The continuous line represents the result of the simulation while dots report an experimental dataset ~\cite{kinzer1951}. The regime temperature of droplet corresponds to the ventilated wet bulb temperature at prescribed environmental pressure and actual temperature, which is $T_{wb}=288.15 K$.}
\label{fig:1}
\end{figure}

We have performed the Direct Numerical Simulation of the evaporation of liquid acetone droplets dispersed within a turbulent air/acetone vapor mixture. 
 The numerical code is constitute by two different modules. An Eulerian algorithm directly evolves the gaseous phase dynamics solving the Low-Mach number formulation of the Navier-Stokes equations \eqref{eq:1}-\eqref{eq:2} (see e.g. \cite{picano2011dynamics,rocco2015curvature} and references therein for validation and tests). A second order central finite differences scheme is adopted on the staggered grid for space discretization, while temporal evolution is performed by a low-storage third order Runge-Kutta scheme. A Lagrangian solver evolves droplets mass, momentum and temperature laws \eqref{eq:3}-\eqref{eq:4}. The temporal integration uses the same Runge-Kutta scheme of the Eulerian phase and second-order accurate polynomial interpolations are used
to calculate Eulerian quantities at droplet positions. We have preliminarily tested the evaporation dynamics in the numerical code in two different test cases. The former case concerns a liquid water droplet carried by a laminar dry jet. In this extremely dilute conditions the Spalding $D^2$ law is a valid analytical solution for the droplet radius evolution over time (Fig. \ref{fig:1:a}). In the latter case, a water droplets freely falling in wet air is considered and the numerical solution for the droplet temperature evolution is compared to an experimental dataset (Fig. \ref{fig:1:b}). 

The computational domain consists of a cylinder extending for $2\pi\times22\,R\times70R$ in the azimuthal, $\theta$, radial, $r$ and axial, $z$, directions. The domain has been discretized by $N_\theta\times N_r\times N_z=128\times225\times640$ points using a non-equispaced, staggered mesh in the radial and axial direction. The mesh has been stretched in order to be of the order of the Kolmogorov length in the downstream evolution. The flow is injected at the center of one base of the cylindrical domain and  streams out at the other end. Time-dependent inflow boundary conditions are prescribed. A fully turbulent velocity is assigned at the jet inflow section (Dirichlet condition) by means of a cross-sectional slice of a fully developed companion pipe flow DNS. The flow is injected through a center orifice while the remaining part of the domain base is impermeable and adiabatic.  A convective  condition is adopted at the outlet and  an adiabatic traction-free condition is prescribed at the side boundary. This side boundary condition makes the entrainment of external fluid possible which
in the present case is dry air.
The gas/vapor mixture is injected at a bulk velocity $U_0=9 m/s$ through a nozzle of radius $R=5 \ 10^{-3} m$. The ambient pressure is set to $p=101300 Pa$ while the initial temperature is fixed to $T=275.15 K$. The injection flow rate of the gas is kept constant fixing a bulk Reynolds number $Re=2 U_0 R/\nu=6000$, with $\nu=1.5 10^{-5} m^2 / s$ the  kinematic viscosity. At the inflow section a near saturation condition is prescribed for the air/acetone vapor mixture, $S=Y_V/Y_{V,s}=0.99$, with $Y_{V,s}$ the vapor mass fraction saturation level at the actual temperature. The acetone mass flow rate is set by the mass flow rate ratio $\Phi_m=\dot m_{act}/\dot m_{air}=0.23$, with $\dot m_{act}=\dot m_{act,L}+\dot m_{act,V}$ the sum of liquid and gaseous acetone mass flow rates and $\dot m_{air}$ the gaseous one. Liquid  monodisperse droplets  with radius $r_{d,0}=6 \mu m$ are injected within the saturated vapor carrier phase. All the droplet characteristics have been chosen to reproduce acetone liquid. The injected droplets are distributed randomly over the inflow section with initial velocity equals to the local turbulent gas phase velocity. Before the injection of droplets, the simulation is started considering only the single-phase flow until statistical steady conditions have been attained. Then the simulation is run for about $200R/U_0$ time scales in order to reach a statistical steady condition for the two-phase evaporating flow before to collect the dataset. The statistics considers around one hundred samples separated in time by $R/U_0=1$.

\section{Results and discussions}

\begin{figure}[t]
\subfigure[]{\includegraphics[width=0.248\textwidth]{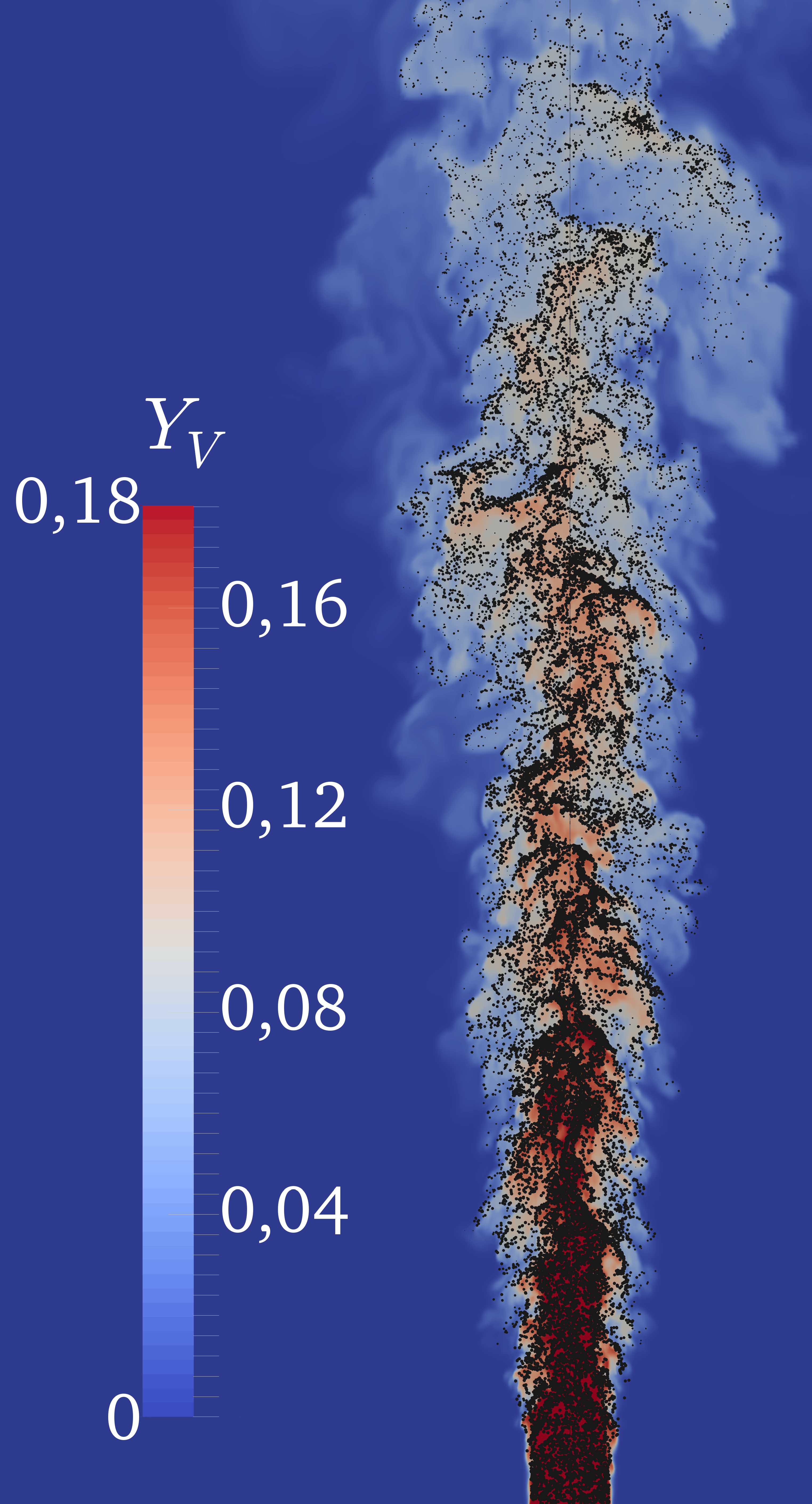} \label{fig:2:a}} \
\subfigure[]{\includegraphics[width=0.198\textwidth]{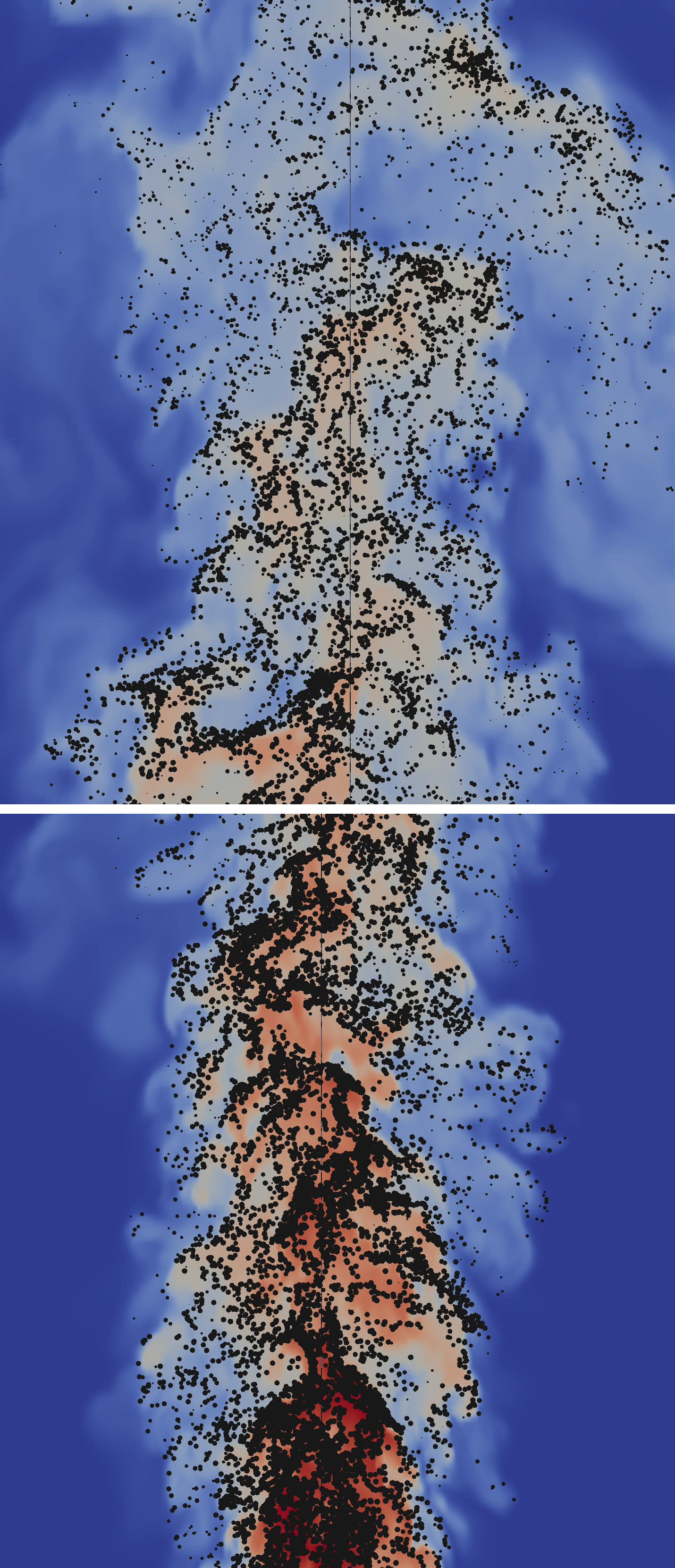} \label{fig:2:b}} \
 \caption{(a) A radial-axial slice of the turbulent spray. The black points represent a subset of droplets formed by 1/5 of the whole population. Only droplets located inside a slice of width $w/R=0.01$ are visible. Each point size is proportional to the corresponding droplet radius (scale factor 100). The carrier phase is contoured according to the instantaneous vapor mass fraction field, $Y_V$, which  is bounded between $0$ and $0.18$, the former corresponding to the dry condition and the latter to the 99\% saturation level prescribed at inlet. (b) Enlargements of two different jet regions centered at $z/R=10$ (lower panel) and $z/R=20$ (upper panel).}
\label{fig:2}
\end{figure}

A general overview of the instantaneous vapor mass fraction field and droplet population distribution is provided in Fig. \ref{fig:2}. Droplets populate only the turbulent jet core, while are not present in the outer dry region. The distribution of droplets is strongly inhomogeneous and clustering is also apparent. In particular, droplets seem to preferentially segregate in regions characterized by high vapor concentration while only few isolated droplets can be found in poorly saturated areas (see the enlargements shown in Fig. \ref{fig:2:b}). There are several mechanisms driving droplets preferential concentration in a turbulent flow, the most relevant of them being the small-scale clustering ~\cite{eaton1994preferential,toschi2009lagrangian}, droplets accumulation along jet axis ~\cite{picano2010anomalous,lau2016effect} and as we will show the droplet dynamics in the mixing layer. The small scale clustering concerns the interaction of droplets distribution with the smallest structures of turbulence that promotes the segregation if the dissipative flow time scale is of the same order of the particle/droplet (inertial) time scale. In turbulent jets or sprays, a mean accumulation of the dispersed phase has been observed at specific distances from the inflow. This location is determined by the matching of particle time scale and the local large scale jet time which quadratically decreases with the downstream axial distance ~\cite{picano2010anomalous,lau2016effect}. 

Independently by the mechanisms driving droplets segregation, the vapor mass fraction increases rapidly inside clusters of evaporating droplets due to their high concentration. As the vapor concentration increases, the local evaporation rate is reduced. The vaporization process may be even completely blocked if the vapor concentration reach the saturation level, $Y_{V,s}$. In this case a non evaporating, fully saturated core appears inside the cluster ~\cite{reveillon2007evaporating}. It is then clear how the clustering phenomenon can strongly affect the overall vaporization process (e.g. evaporation length) by reducing locally the rate of vaporization. 

\begin{figure}[t]
\subfigure[]{\includegraphics[width=0.45\columnwidth]{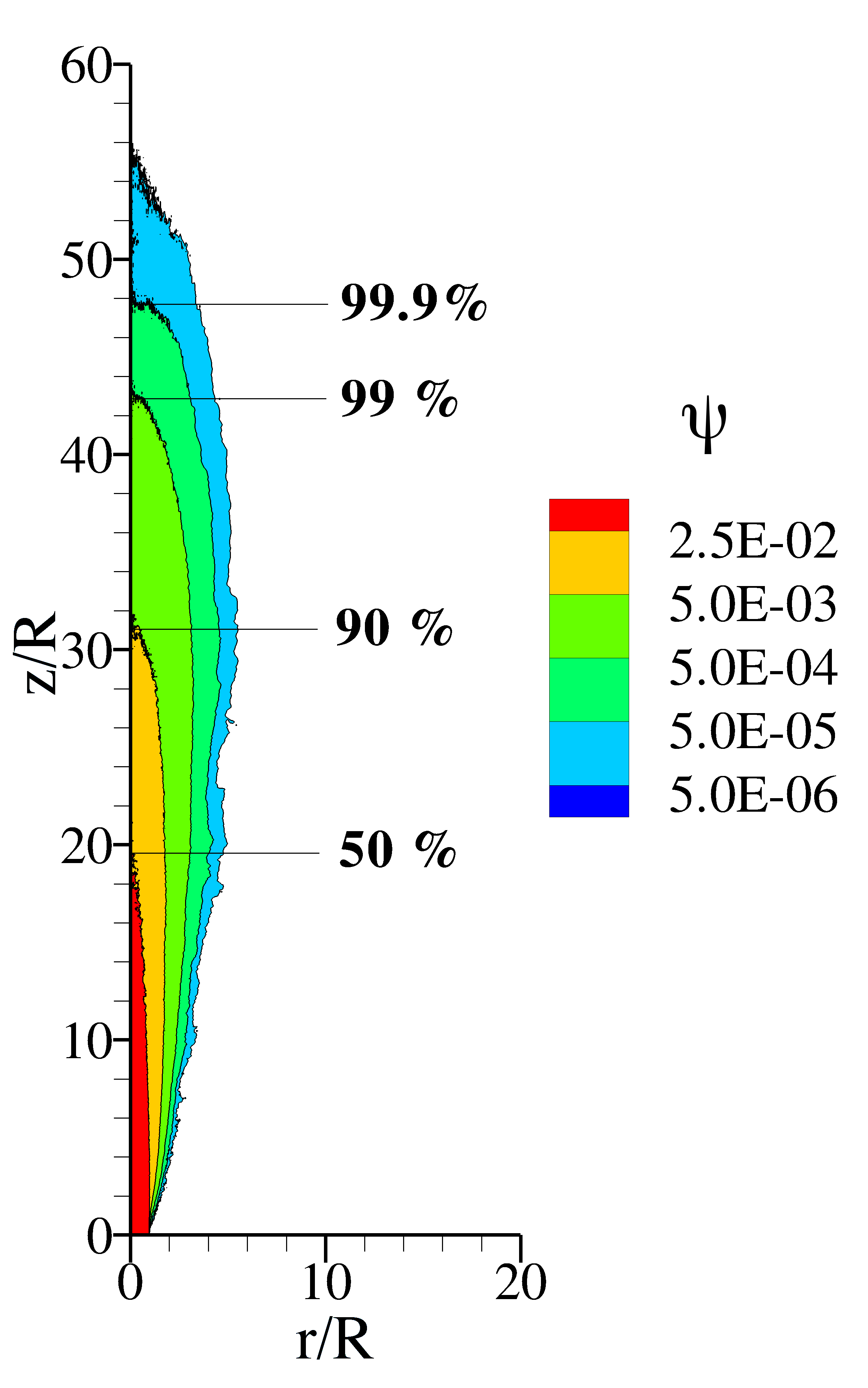} \label{fig:3:a}} \
\subfigure[]{\includegraphics[width=0.45\columnwidth]{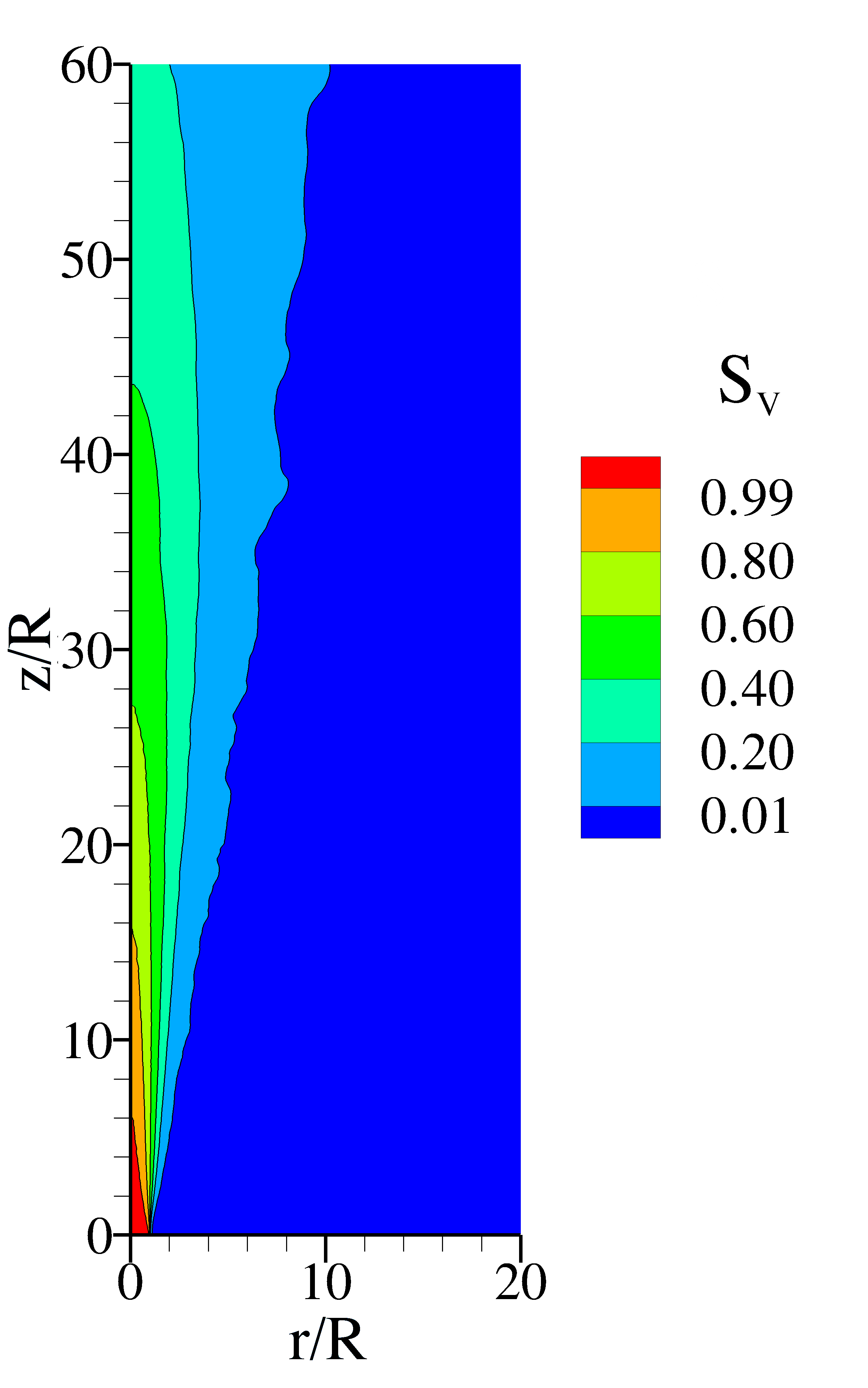} \label{fig:3:b}} \
\caption{(a) Average liquid phase mass fraction, $\Psi=m_L/m_{gas}$ where $m_L$ is the mean mass of liquid acetone and $m_{G}$ is the mean mass of the gaseous carrier phase.
The labels show different distances from the jet inlet, $z/R$, in correspondence of which the 50\%, 90\%, 99\% and 99.9\% of the injected liquid mass is evaporated. (b) Average saturation field, $S=Y_{V}/Y_{V,s}$, where $Y_V$ is the actual vapor mass fraction field and $Y_{V,s}=Y_{V,s}(p,T)$ is the value of vapor mass fraction corresponding to the local saturation condition.}
\label{fig:3}
\end{figure}

The Fig. \ref{fig:3:a} provides the average distribution of liquid mass fraction over the spray, $\Psi=m_L/m_{G}$, where $m_L$ is the mean mass of liquid acetone and $m_{G}$ is the mean mass of the gaseous carrier phase inside an arbitrary small control volume, $\Delta V$. The overall vaporization length can be defined as the axial distance from inlet where the 99$\%$ of the injected liquid mass has transit to vapor phase. According to this definition, the vaporization process is terminated at about $z/R\simeq43$. It should be noted that the liquid phase mass fraction is significantly higher in the spray core then in the outer layer, such that the turbulent gaseous phase dynamics is especially affected by the presence of the dispersed phase in this region. This is consistent with the distribution of the average saturation field, $S$, reported in Fig. \ref{fig:3:b}. We note an almost-saturated flow near the inlet caused by the prescribed inflow conditions, then the saturation level gently decreases in the downstream evolution while sharply towards the outer jet region. The turbulent spray is constituted by a spreading and slowly decaying turbulent core which is surrounded by the dry and irrotational environmental air. The turbulent core is continuously entrained by the environmental air which mixes up with the turbulent air/vapor mixture thus reducing the vapor concentration. Since the inner core fluid cannot reach the outer region, the spray core shows higher saturation level over the whole downstream evolution of the flow. The effect of dry air entrainment is  crucial on the overall vaporization process. The dilution of vapor concentration is indeed fundamental in order to allow the vaporization process to advance.

\begin{figure}[t]
\subfigure[]{\includegraphics[width=0.45\columnwidth]{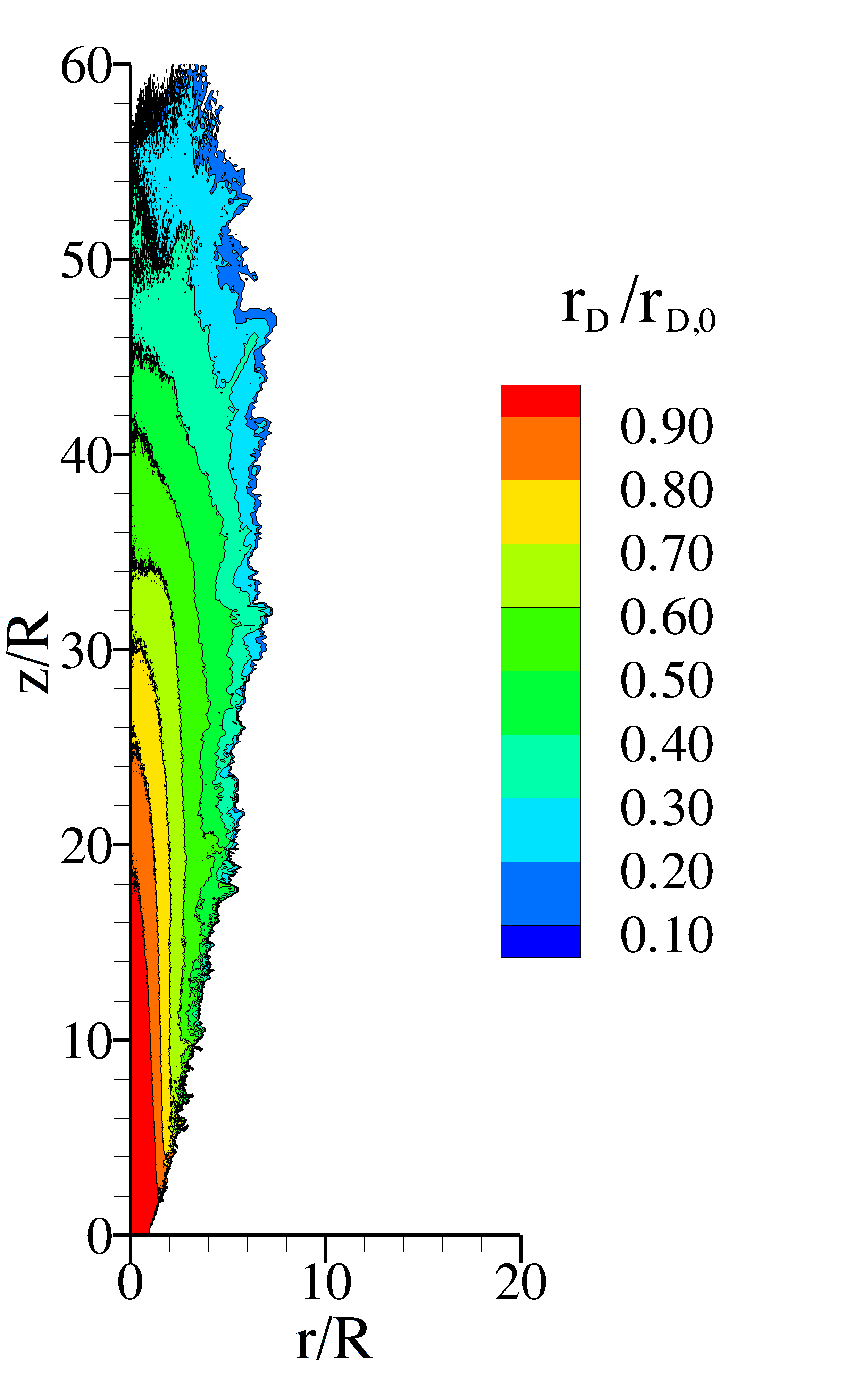} \label{fig:4:a}} \
\subfigure[]{\includegraphics[width=0.45\columnwidth]{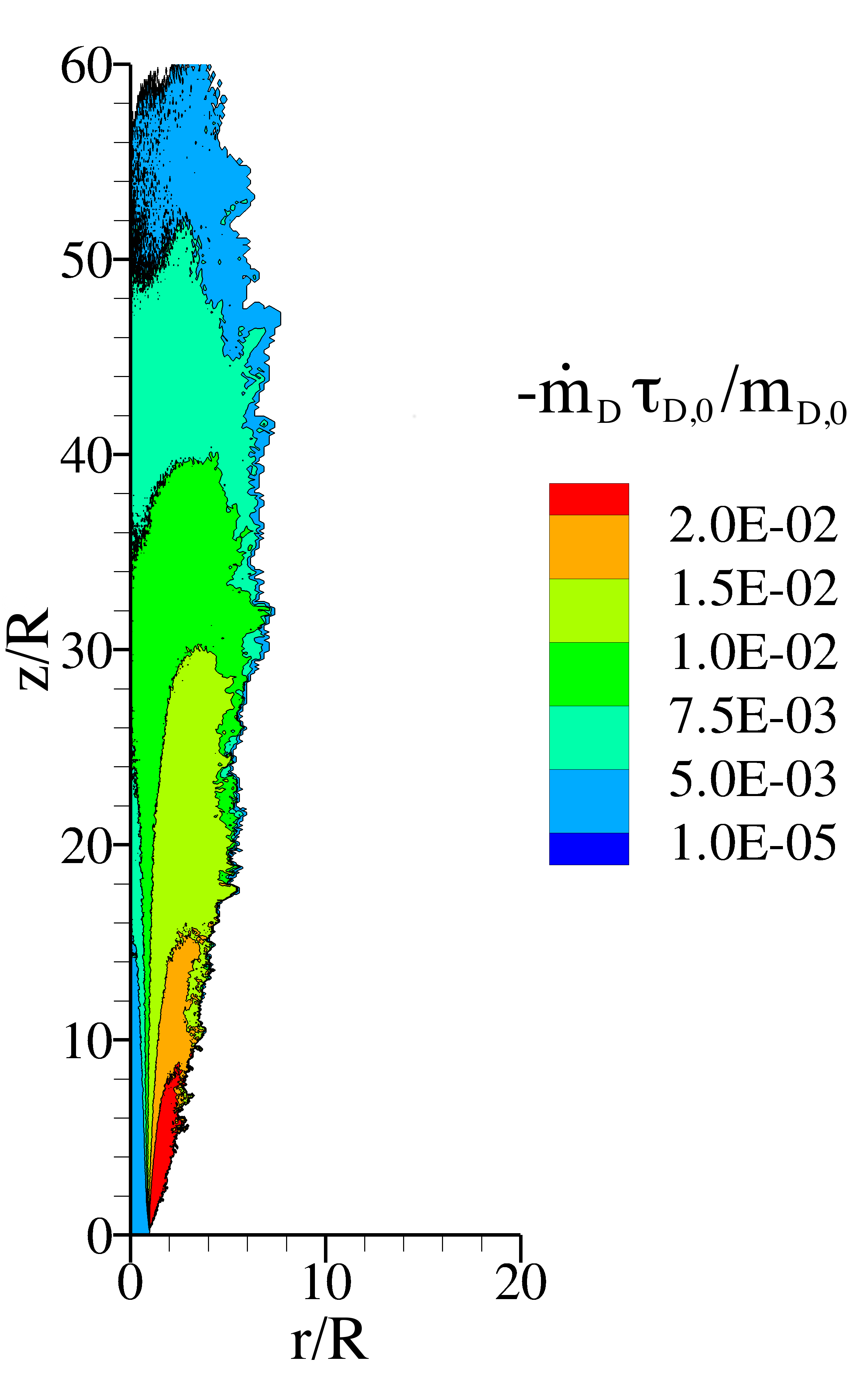} \label{fig:4:b}} \
\caption{(a) Average droplet radius rescaled by the initial droplets radius $r_{d,0}$. (b) Average droplet vaporization rate divided by the reference scale defined as $\dot{m_{d,0}}=m_{d,0}/\tau_{d,0}$  with $m_{d,0}$ the initial droplet mass and $\tau_{d,0}$ the initial droplet relaxation time.}
\label{fig:4}
\end{figure}

The average droplets radius and evaporation rate distributions are reported respectively in Fig. \ref{fig:4:a} and Fig. \ref{fig:4:b}. According to the discussed entrainment effect, the vaporization rate is maximum in the mixing layer separating the jet outer and core regions. The peak value is reached in the shear layer immediately downstream the inflow section, where large droplets enter in direct contact with the dry environmental air. Consequently, at each axial distance form inlet, larger droplets can be found in the spray core where the vaporization process is slowed down by the high vapor concentration, while smaller droplets can be found towards the outer region where the vaporization proceeds faster.

\begin{figure}[t]
\subfigure[]{\includegraphics[width=0.45\columnwidth]{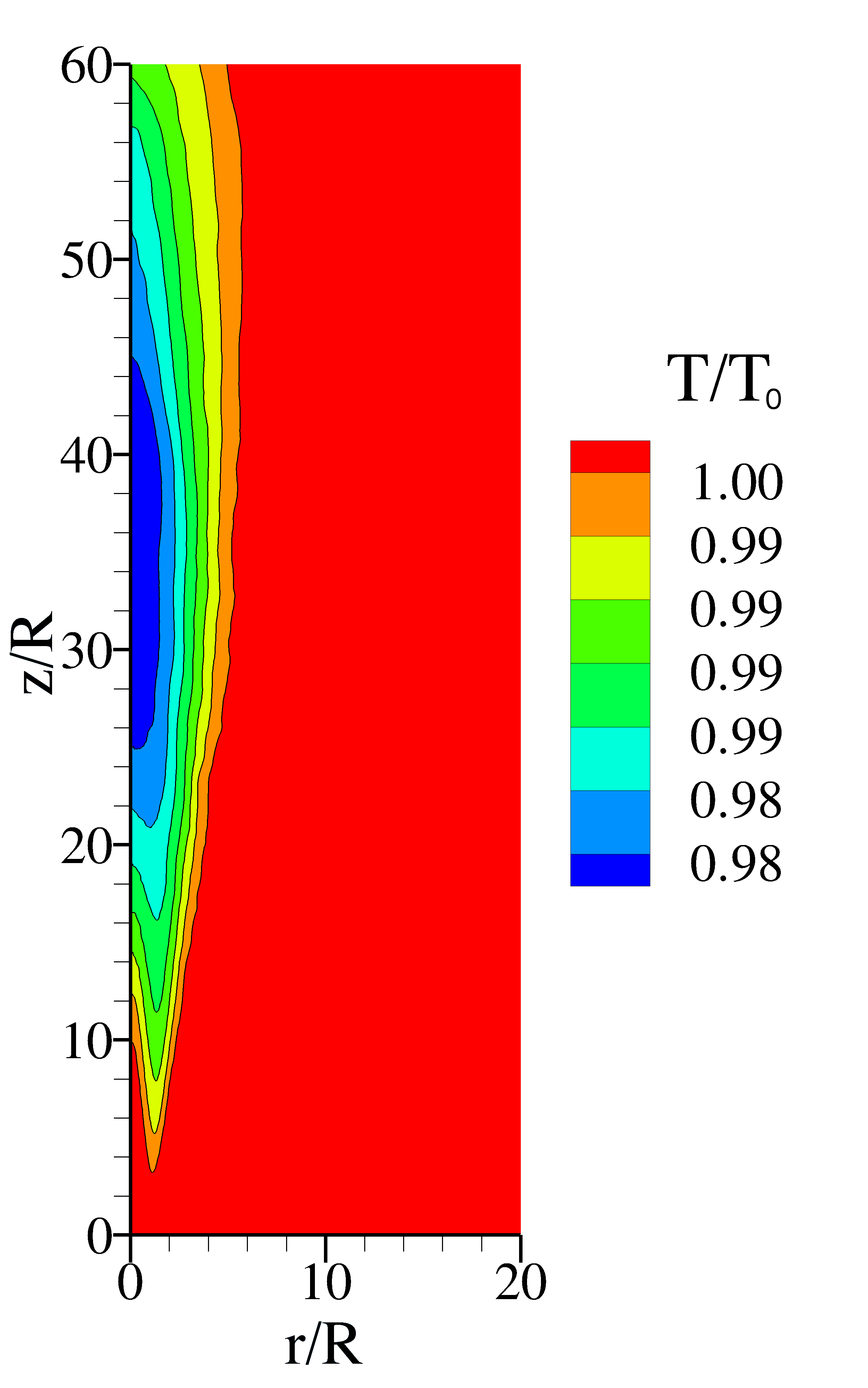} \label{fig:5:a}} \
\subfigure[]{\includegraphics[width=0.45\columnwidth]{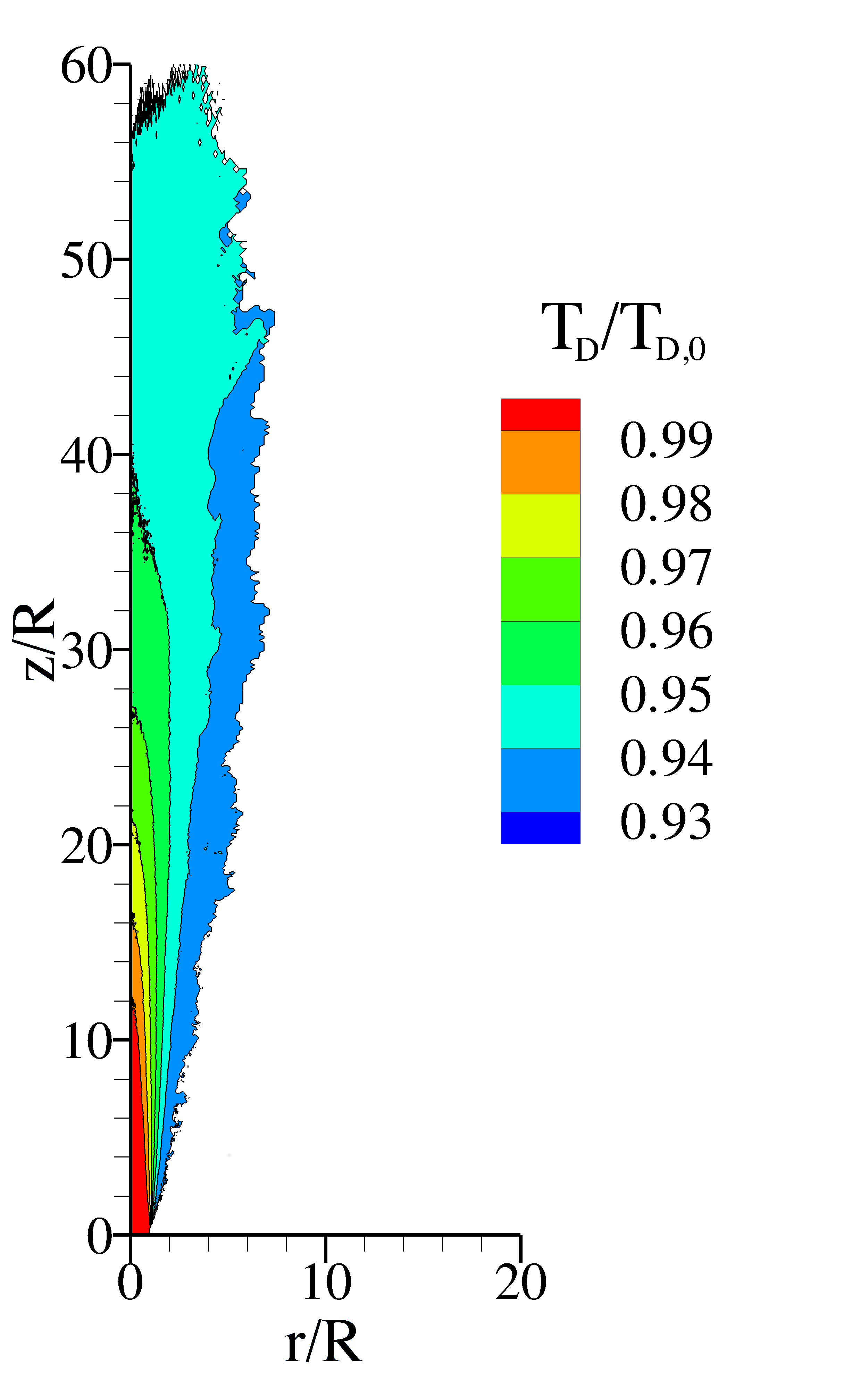} \label{fig:5:b}} \
\caption{(a) Average gas phase temperature, $T$, rescaled by the injection temperature, $T_0$. (b) Average droplet temperature, $T_d$ rescaled by droplet initial temperature, $T_{d,0}$. The reference temperature scales are equal $T_d=T_{d,0}$.}
\label{fig:5}
\end{figure}

The transition of liquid phase to vapor phase requires an amount of energy per unit mass equal to the latent heat of vaporization of acetone so that the power required by the vaporization is proportional to the evaporation rate. The overall energy required by the vaporization process is provided by the internal energy of both the gaseous carrier phase and the liquid dispersed phase, thus resulting in an overall cooling of the spray in the downstream evolution. The average distribution of the carrier phase and the droplets temperature is reported in Fig. \ref{fig:5}. In the outer spray region the smaller droplets surrounded by low-saturated gas are colder than core droplets due to the higher evaporation rate. Nevertheless, the carrier phase shows an opposite behavior: the spray core is sensibly colder then the shear layer and the minimum gas temperature can be observed between $z/R=25$ and $z/R=45$. This phenomenon is due to the distribution of the liquid phase mass fraction. In the spray core the liquid mass represent a significant part of the overall spray mass. Hence, the cooling effect due to vaporization is much more intense in this region where a large amount of droplets slowly evaporate.

\begin{figure}[t]
\subfigure[]{\includegraphics[width=0.45\columnwidth]{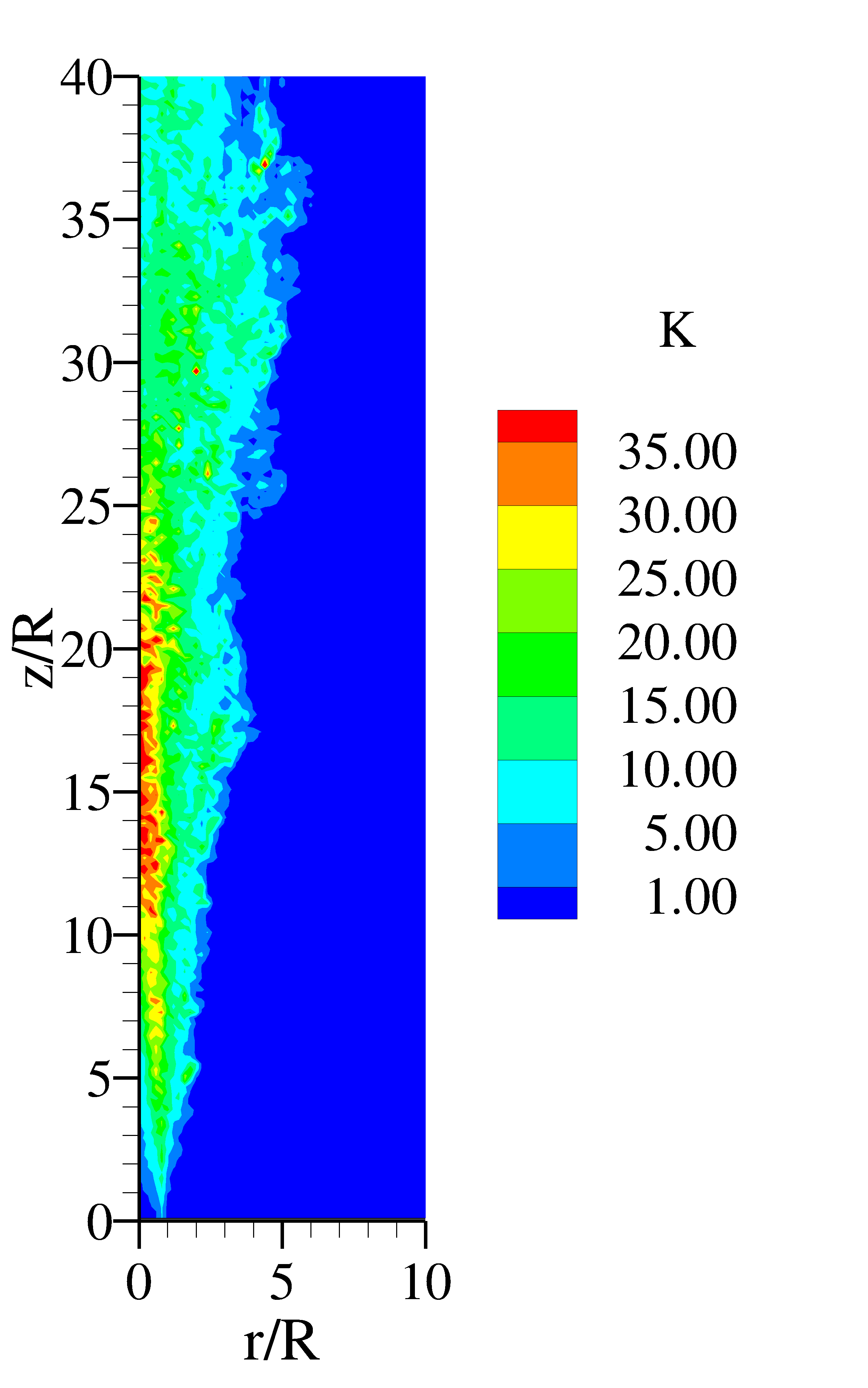} \label{fig:6:a}} \
\subfigure[]{\includegraphics[width=0.50\columnwidth]{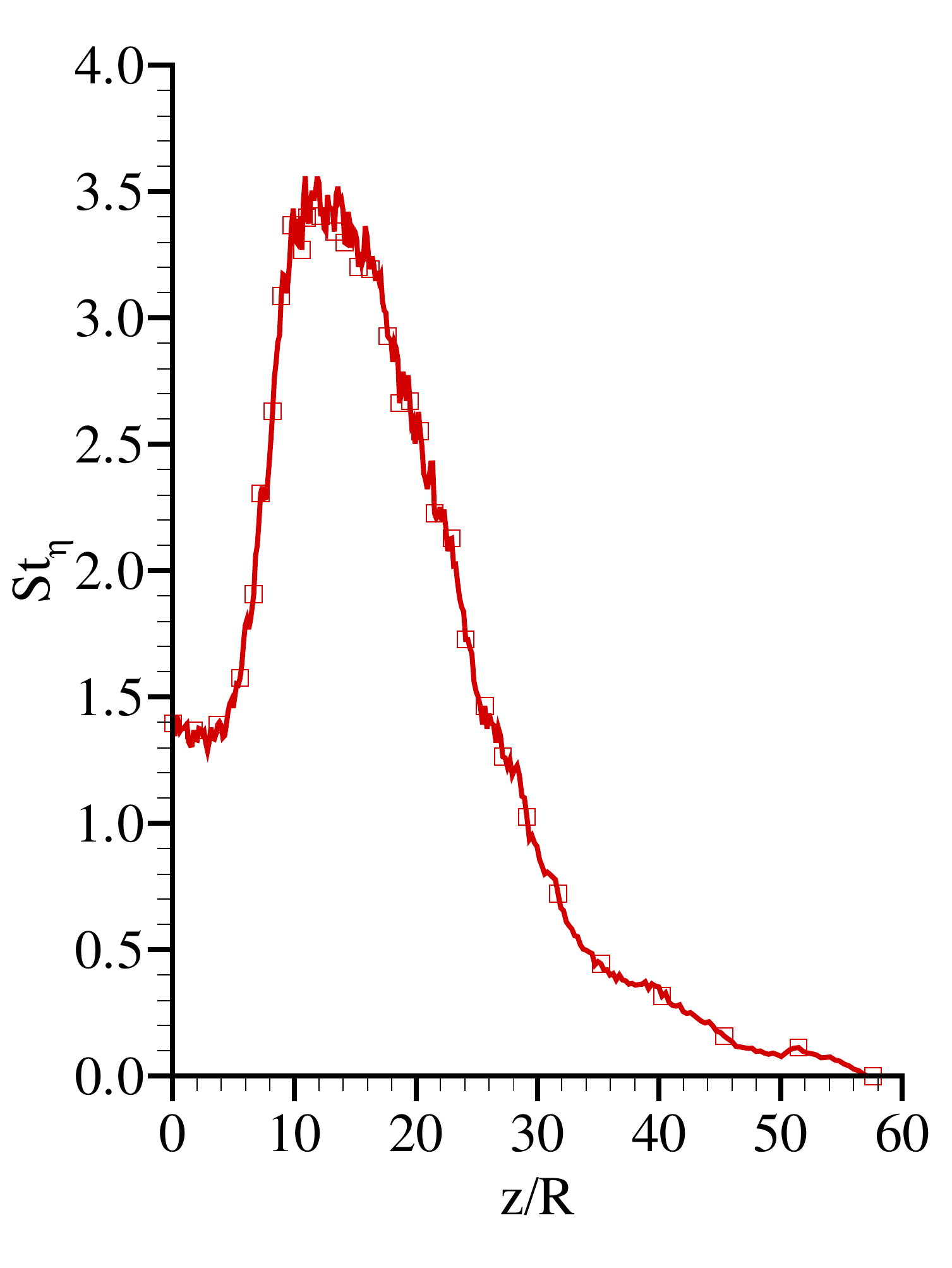} \label{fig:6:b}}
\caption{(a) Droplet clustering index, $K$. 
(b) Evolution of the mean droplets Stokes number, $St_\eta=\tau_d/\tau_\eta$, based on the Kolmogorov dissipative scales on the jet axis. 
}
\label{fig:6}
\end{figure}

We have highlighted the existence of a strong preferential segregation of droplets, focusing in particular on the effect of this inhomogeneous distribution on the overall vaporization process dynamics. The intensity of droplets segregation can be measured in different ways, e.g.\cite{shaw2002towards}. 
We will measure the intensity of clustering in each point of the inhomogeneous turbulent jet spray\cite{battista2011intermittent} using the clustering index $K$, which is defined as

\begin{equation}
K=\frac{\overline{(\delta n)^2}}{\overline{n}}-1
\end{equation}

where $\overline n$ and $\overline{(\delta n)^2}$ are the mean and variance of the number of droplets in given small volume $\Delta V$. If droplets are
completely randomly located, their distribution is determined by a Poisson process in which mean and variance coincide. Hence if clustering is not present
and particles are random distributed $K=0$. 
%
%
On the opposite, if $K > 0$  the variance exceeds the mean value indicating that droplets preferentially segregate in clusters. Fig. \ref{fig:6:a} shows the clustering index $K$ computed over the whole spray domain. The large positive value of $K$ corresponds  to strong preferential segregation of droplets. 
We note that droplets are initially random distributed and then develop clustering. In particular, near the inflow $K$ assumes positive value only in the mixing 
layer where the local droplet concentration is intermittent because of the fluctuation of the turbulent/non-turbulent interface which separates the turbulent core populated by droplets and the outer region without droplets. 
It should be noted that the air regions entrained from the environment in the core are almost droplet-free and enhance the fluctuation level of 
the droplet concentration even in the jet core, see the snapshots reported in Fig~\ref{fig:2}. 
Downstream the clustering appears in the whole turbulent jet core. We attribute this phenomenon also to  the developing of small-scale turbulent clustering.  

The main mechanism driving the small scale clustering relies in the competition between inertia and Stokes drag. The drag tends to trail droplets according to the highly convoluted local turbulent structures while droplets finite inertia prevents them to follow exactly the turbulent flow motion. By this mechanism droplets heavier then the fluid tend to be ejected from vortex cores ~\cite{bec2007heavy}. The small-scale droplet distribution is governed by the Stokes number, $St_\eta$ , which is defined as the ratio of droplets response time, $\tau_d=2 \rho_L r_d^2/(9 \rho \nu)$, and the characteristic time of the dissipative scales, $\tau_\eta=(\nu/\varepsilon)^{1/2}$. Droplets with $St_\eta \gg 1$ act as ballistic particles that move across turbulent structures being only weakly perturbed and showing a negligible preferential segregation. On the opposite, droplets with  $St_\eta \ll1$ act as passive tracers which move according to the local turbulent motion without exhibiting clustering. Preferential segregation is maximum when $St_\eta \sim 1$. 
The Fig. \ref{fig:6:b} provides the Stokes number, $St_\eta$, of droplets located within a radial distance $r/R=0.2$ from the jet centerline. In this region the small-scale clustering should be the most significant with respect to the preferential segregation effect induced by the intermittency of the external shear layer.  The Stokes number decreases in the downstream evolution of the spray assuming unity value around $z/R\simeq25$.
Hence it appears that droplets show an intense clustering promoted both by the small-scale turbulent clustering and by the entrainment process associated to the fluctuation of the interface which separates the inner core region rich of droplet and the outer region which is depleted of droplets.

\begin{figure}[ht]
\subfigure[]{\includegraphics[width=0.48\columnwidth]{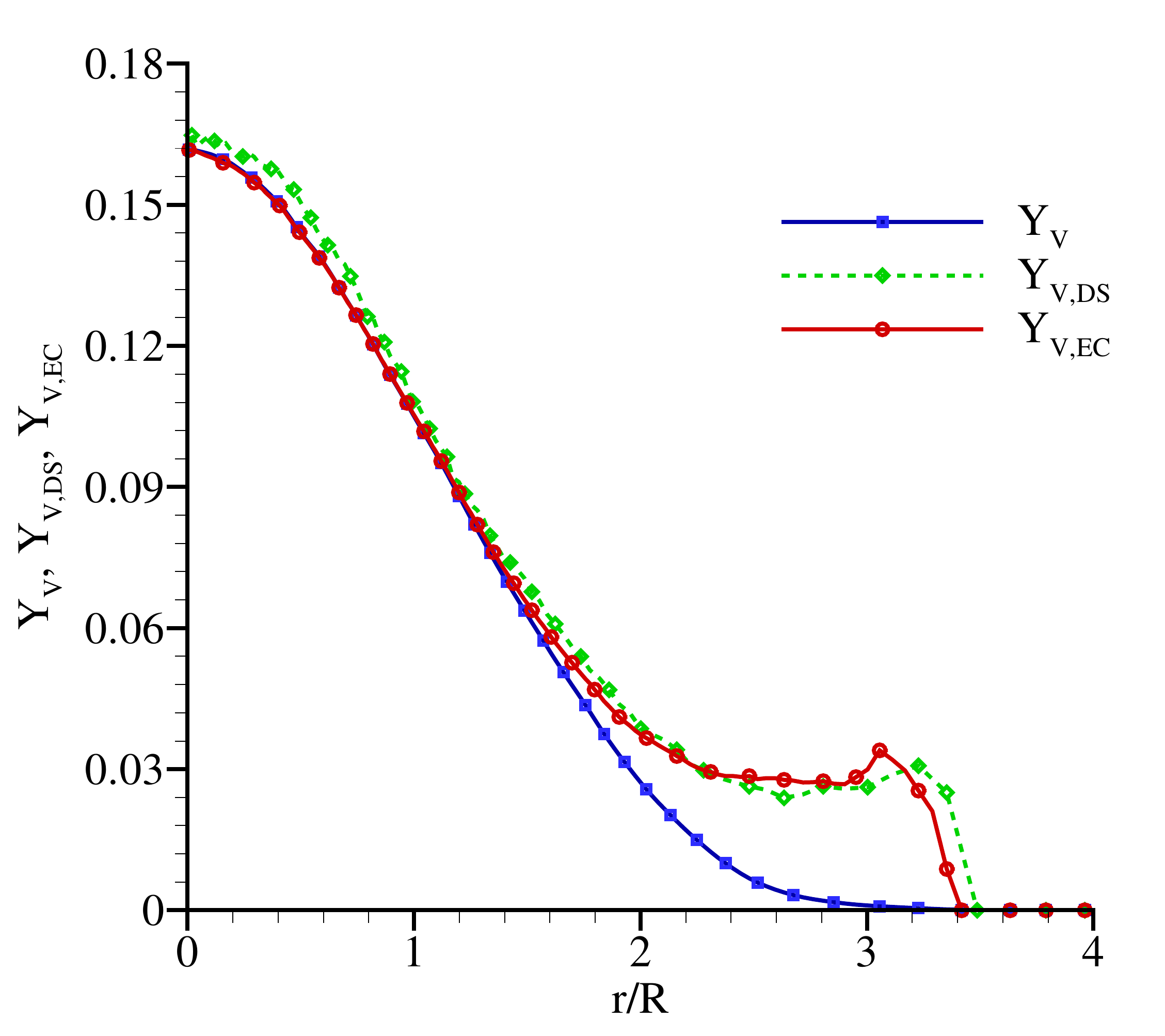} \label{fig:9:a}} 
\subfigure[]{\includegraphics[width=0.48\columnwidth]{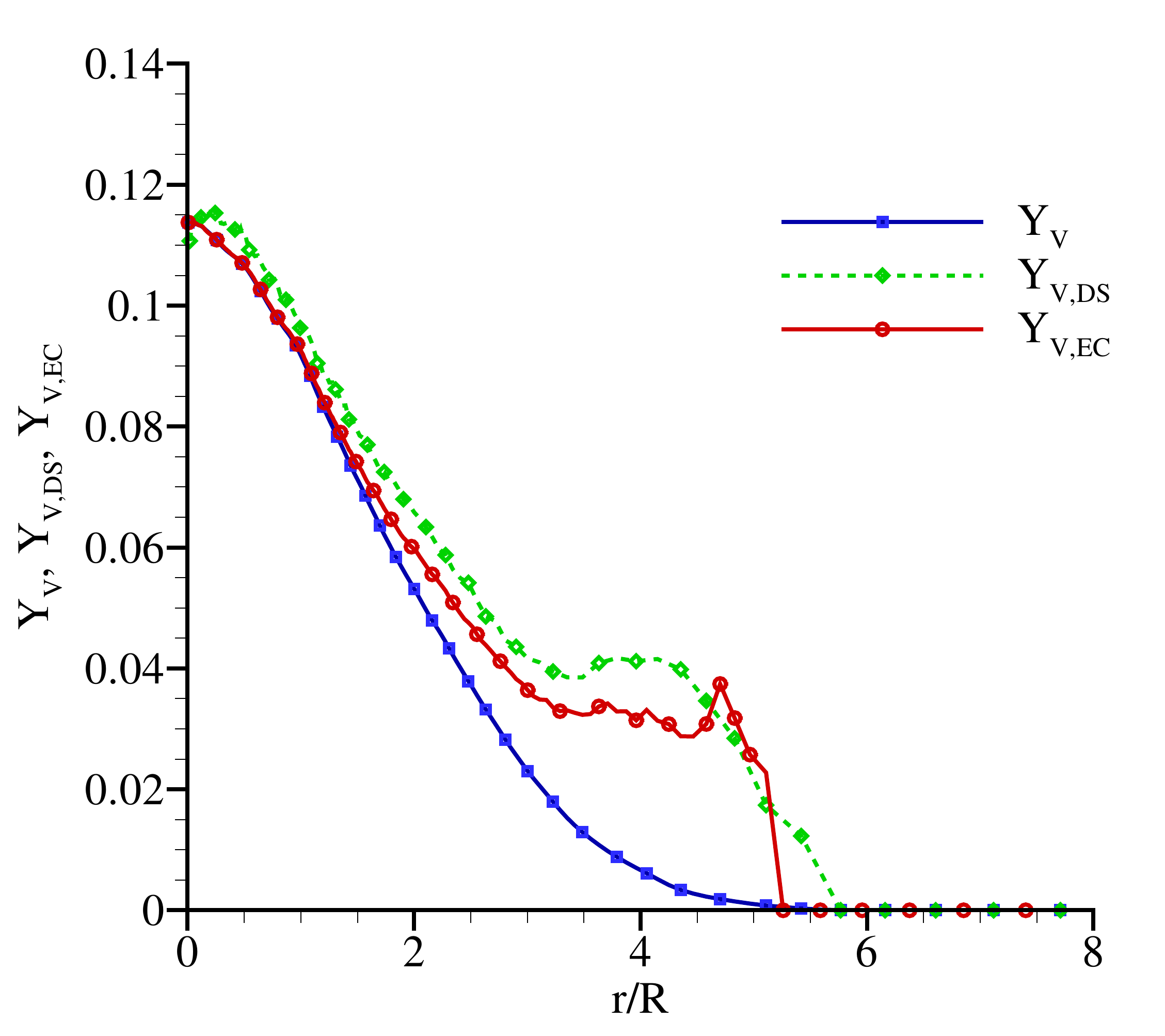} \label{fig:9:b}}\\
\subfigure[]{\includegraphics[width=0.48\columnwidth]{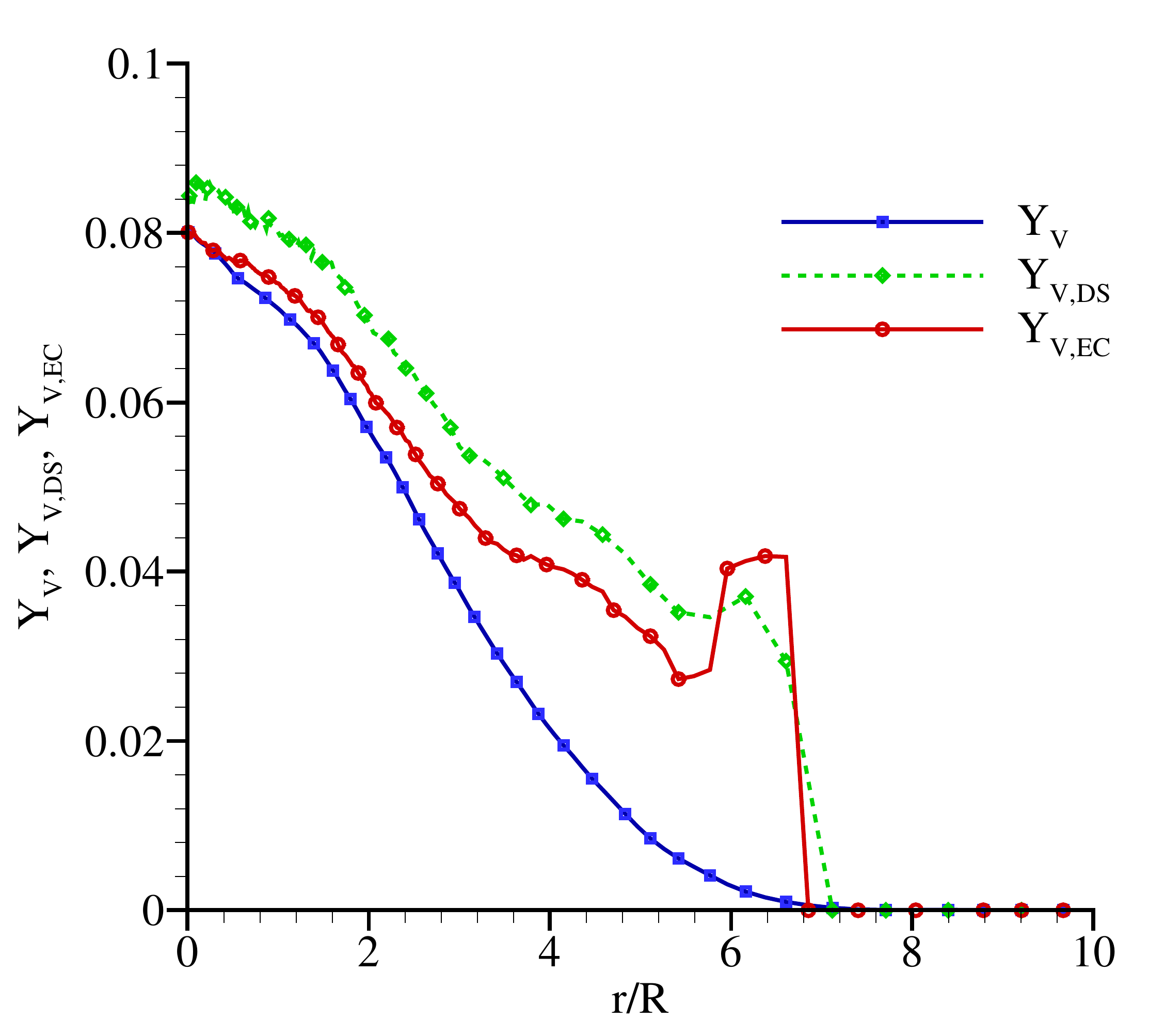} \label{fig:9:c}} 
\subfigure[]{\includegraphics[width=0.48\columnwidth]{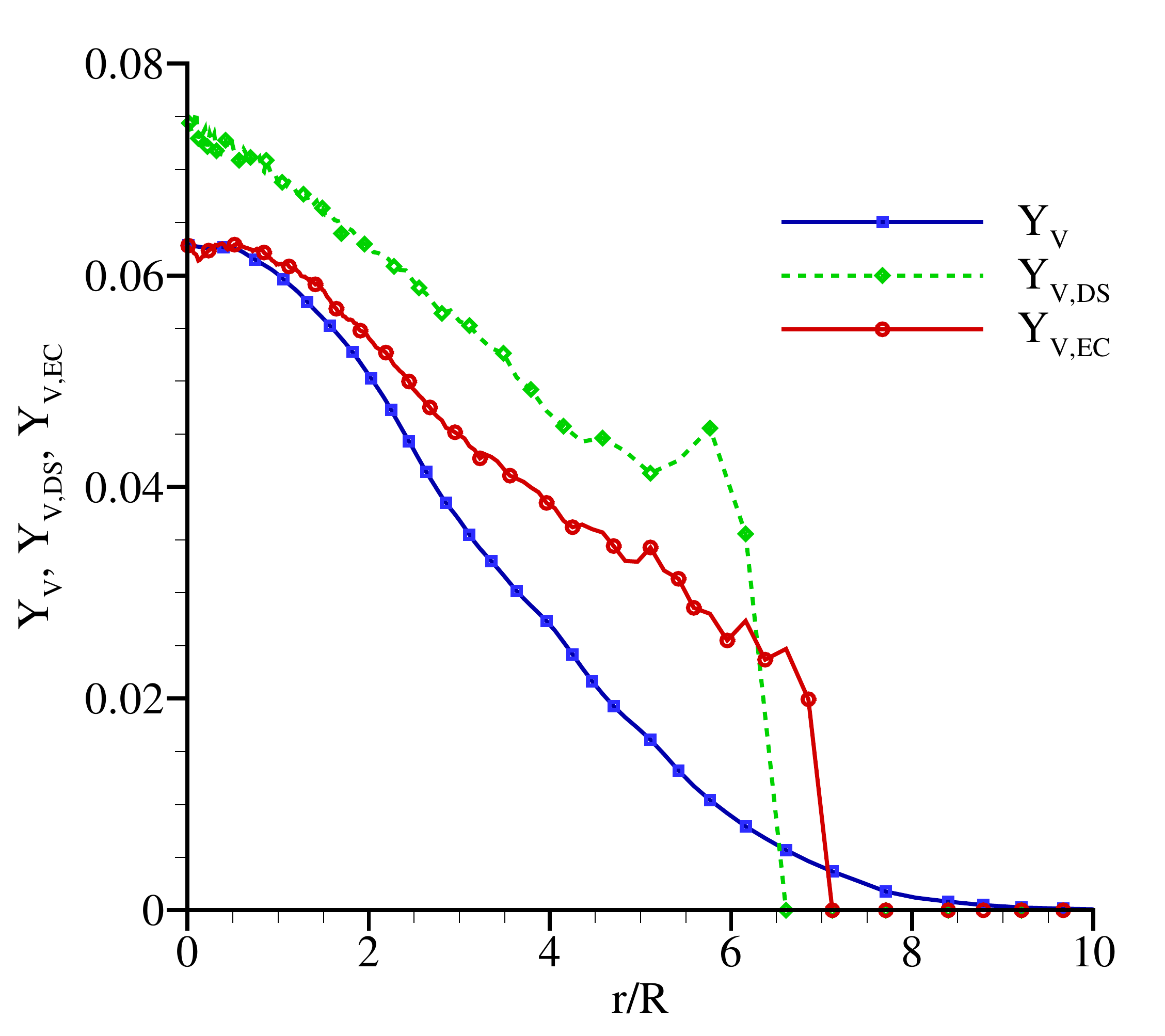} \label{fig:9:d}}
\caption{The figure report the radial profiles of the average vapor mass fraction field at four different axial distances from the origin: (a) $z/R=10$, (b) $z/R=20$, (c) $z/R=30$ and (d) $z/R=40$. Each plot shows the enstrophy-threshold conditional average, $Y_{v,ec}$, the droplet-presence conditional average, $Y_{v,dc}$ and the unconditional Eulerian one, $Y_v$. The enstrophy-threshold conditional average is calculated by sampling the vapor mass fraction only over turbulent core events ($I=1$), that is when local enstrophy exceeds a fixed threshold. $Y_{v,dc}$ is the vapor concentration field obtained by a conditional average on the droplet presence in a given point.}
\label{fig:9}
\end{figure}

To quantify the importance of the droplet clustering in the evaporation process we compare the mean vapor concentration field \emph{felt} by the droplets, 
$Y_{V,dc}$, and the unconditioned Eulerian one, $Y_V$. $Y_{V,dc}$ is the vapor concentration field obtained by a conditional average on the droplet presence in a given point. Hence if droplets preferentially accumulate in locations where the vapor concentration is relatively high it results $Y_{V,dc}>Y_V$. 
Figure~\ref{fig:9} reports the radial profiles of  $Y_{V,dc}$ and $Y_V$ at different axial distances $z/R$. 

The vapor mass fraction felt by droplets is usually higher then the correspondent unconditional value. At $z/R=10$ the droplet conditioned and unconditioned
vapor concentration are similar with the exception of the outer part. This behaviour is expected since we have observed the clustering to be small 
near the inflow with the exception of the mixing layer. The clustering associated to the mixing layer, separating the outer and inner jet regions, 
will be discussed in details in the following. At higher $z/R$ the preferential sampling of the vapor phase operated by segregating droplet is significant  
with an oversampling of about $10\div40\%$ more the unconditioned value even in the inner jet core. 
To characterize droplet dynamics in the mixing layer, we need to discern between the inner turbulent jet core and the irrotational outer region.
The two regions are separated by an almost sharp fluctuating layer, so-called turbulent/non-turbulent interface~\cite{da2014interfacial}, that is highly convoluted over a wide range of turbulent scales. The most used observable to characterize the two regions is 
 the local enstrophy, $\zeta^2=  || \nabla \times \vec u || ^2$.
The inner turbulent core is characterized by large fluctuations of enstrophy while in the outer region the enstrophy is null.  Thus, fixing an entrophy threshold, $\zeta^2_{th}$, it is possible to distinguish if a point is located into the turbulent region or not:

\begin{equation}
I(\vec x, t)=H[\zeta^2(\vec x, t) -\zeta^2_{th}]
\end{equation}
 
\noindent with $H$ the heaviside function. $I=1$ denotes a turbulent event, while $I=0$ an irrotational one. 
The value of $\zeta_{th}$ has been shown to play a 
weak influence~\cite{da2014interfacial}. Using $I$ we can define an enstrophy-threshold conditional average for the vapor concentration, 
$Y_{V,ec}$, by sampling the vapor mass fraction only over turbulent core events. Besides the unconditioned and the droplet conditioned statistics, 
Fig. \ref{fig:9} provides also $Y_{V,ec}$ which 
can be seen as the mean concentration field of the turbulent core region. In the mixing layer, this turbulent conditional average shows an excellent agreement
with the vapor concentration felt by the droplets which is significantly higher than the unconditioned value. Since the unconditioned value $Y_V$ is determined 
both by irrotational dry outer and turbulent vapor-concentrated events, the present analysis indicates that the droplet dynamics in the mixing layer is mainly 
determined by turbulent events. 
In other words, in the mixing layer, 
droplets moving towards the outer region from  the inner turbulent core are surrounded by highly concentrated vapor gas ejected with the droplets. On the contrary, in a point of the mixing
 layer, a low vapor concentration event is associated to an engulfment of entrained dry air which is depleted of droplets. 
 Because of this dynamics, in average, droplets
 evaporating in the mixing layer do not feel the unconditioned mean vapor concentration, but a higher level.     
At further downstream distance the effect
of small-scale clustering previously discussed adds its contribution to this dynamics.

Hence, since droplet vaporization rate is driven by the vapor concentration sampled by droplets, the oversampling of the vapor concentration field slows down the overall vaporization process, thus increasing the overall vaporization length and time. We find that droplets preferential sampling is the results of two different contributions, originated by distinct mechanisms. The first mechanism is induced by the fluctuation of the turbulent/non-turbulent interface in the mixing layer. In this area droplets are entrapped in turbulent structures with high vapor concentration that protract into the irrotational, droplet-free ambient gas. 
The second contribution is given 
by the inertial small scale clustering. Clusters of evaporating droplets move together with their own highly saturated atmosphere induce an oversapling of
the vapor concentration with respect to the neighboring environment. 
 The contribution of this latter mechanism is more evident in the spray core and tend to increase in intensity in the downstream evolution of the spray, while
 the former is dominant in the mixing layer.

\subsection {Probability density function of droplet observables}
 
\begin{figure}[ht]
\centering
\subfigure[]{\includegraphics[width=0.47\columnwidth]{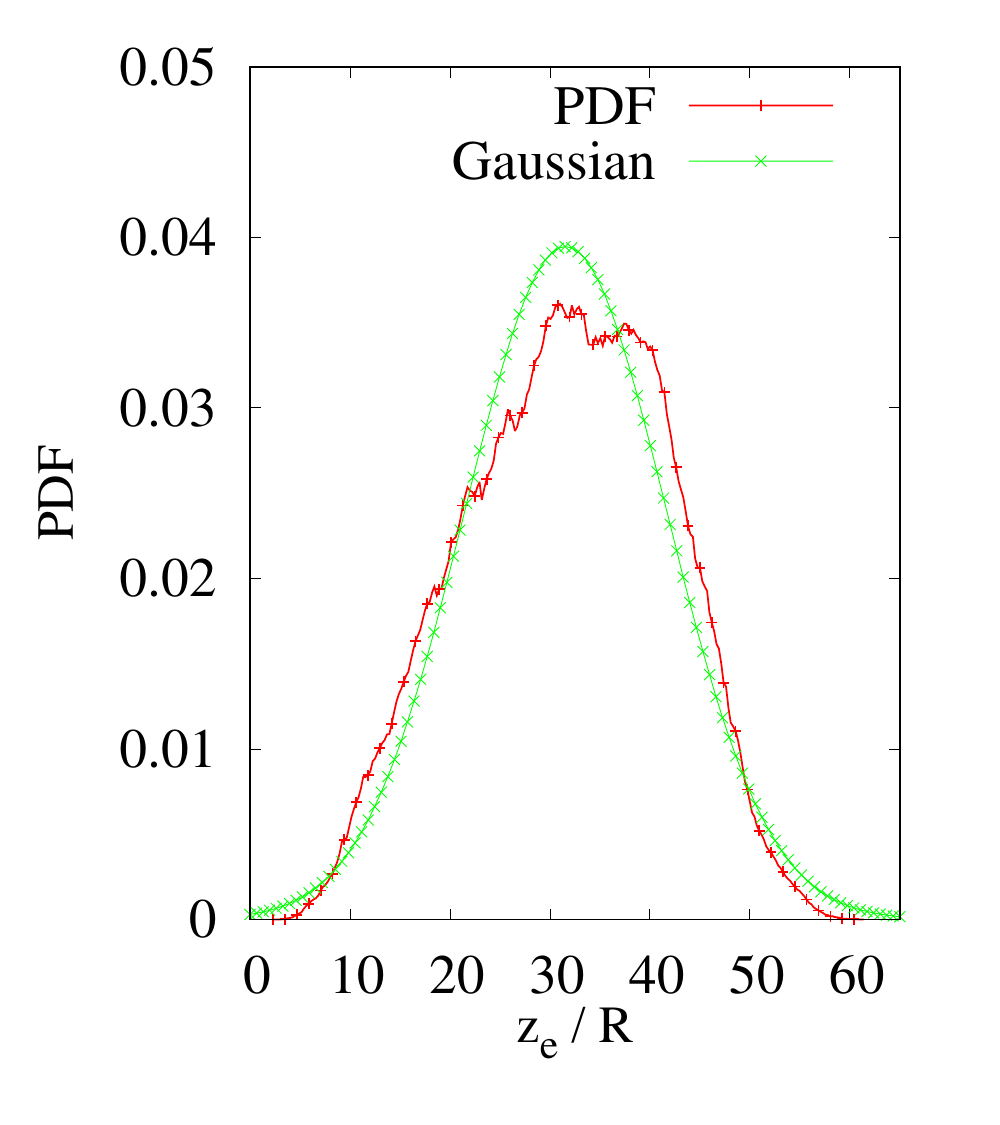} \label{fig:7:a}} \
\subfigure[]{\includegraphics[width=0.47\columnwidth]{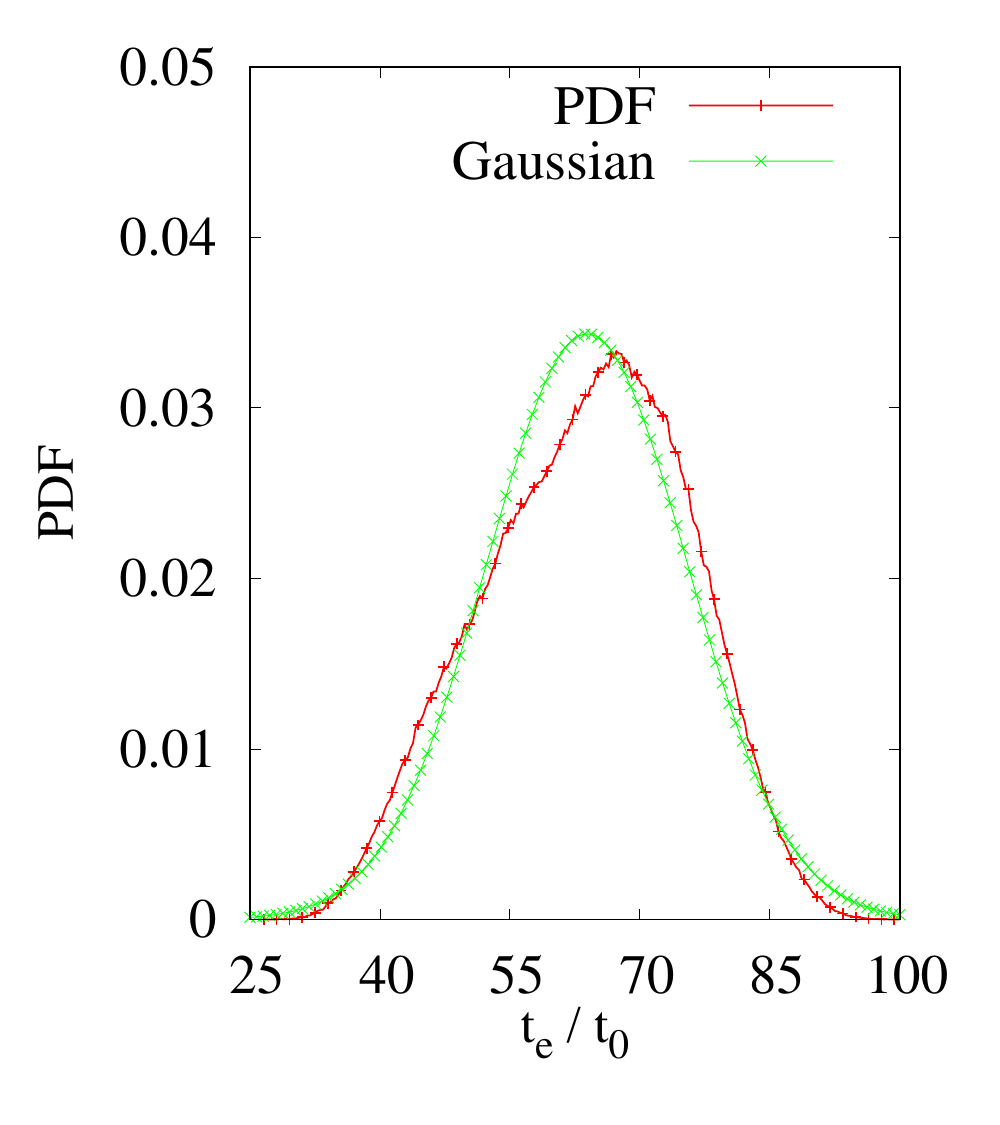} \label{fig:7:b}} \
\caption{ (a) Probability density function of non-dimensional droplet vaporization length, $z_e/R$, with R the jet inlet radius. (b) Probability density function of non-dimensional droplet vaporization time, $t_e/t_0$, with $t_0$ the reference time scale, $t_0=R/U_b$. The PDFs are computed over the entire droplets population injected into the computational domain.}
\label{fig:7}
\end{figure}

In order to further characterize droplet vaporization dynamics, we consider the probability density function of the vaporization length and time computed over the whole droplets population. In analogy with the overall vaporization length definition, one single droplet vaporization length can be defined as the axial distance from inlet, $z_e$, necessary for the the droplets radius to decrease from $r_{d,0}$ to a threshold radius $r_{d,th}=.01 r_{d,0}$. The vaporization time $t_e$ is the corresponding amount of time. The PDFs for droplets evaporation length and time are reported in Fig. \ref{fig:7:a} and \ref{fig:7:b} respectively. The mean, standard deviation, skewness and kurtosis are reported in table \ref{tab:1} and show a nearly Gaussian behavior with significant standard deviations. The Gaussian behavior in turbulent flows is usually associated to fluctuations induced by the large-scale motions.  It is remarkable how different are the histories of the droplets: half of the injected droplets is still present at about $z/R\simeq32$ where about $90\%$ of the injected liquid mass fraction is evaporated. This aspect is connected to the high polydispersity developed by the droplets.

\begin{table}[h]
\centering
\begin{tabular}{|l|c|c|c|c|}
\hline
                            & $\mu$  & $\sigma$ & $K$  & $S$  \\ \hline
\ \ $z_e/R$   \ \    & 30.64 & 10.22 & 2.39 & -0.010      \\ \hline
\ \ $t_e/t_0$  \ \   & 60.22 & 14.06 & 2.71 & -0.378      \\ \hline
\end{tabular}
\label{tab:1}
\caption{The table provides the mean, $\mu$, standard deviation, $\sigma$ , kurtosis, $K$ and skewness, $S$, of the PDFs of droplets evaporation length and time with $\mu=E[X]$,  $\sigma=\sqrt{E[(X-\mu)^2]}$, $K=E[(X-\mu)^4]/E[(X-\mu)^2]^2$ and $S=E[(X-\mu)^3]/E[(X-\mu)^2]^{3/2}$. All variables are non-dimensional.}
\end{table}

\begin{figure*}[ht]
\subfigure[]{\includegraphics[width=0.45\textwidth]{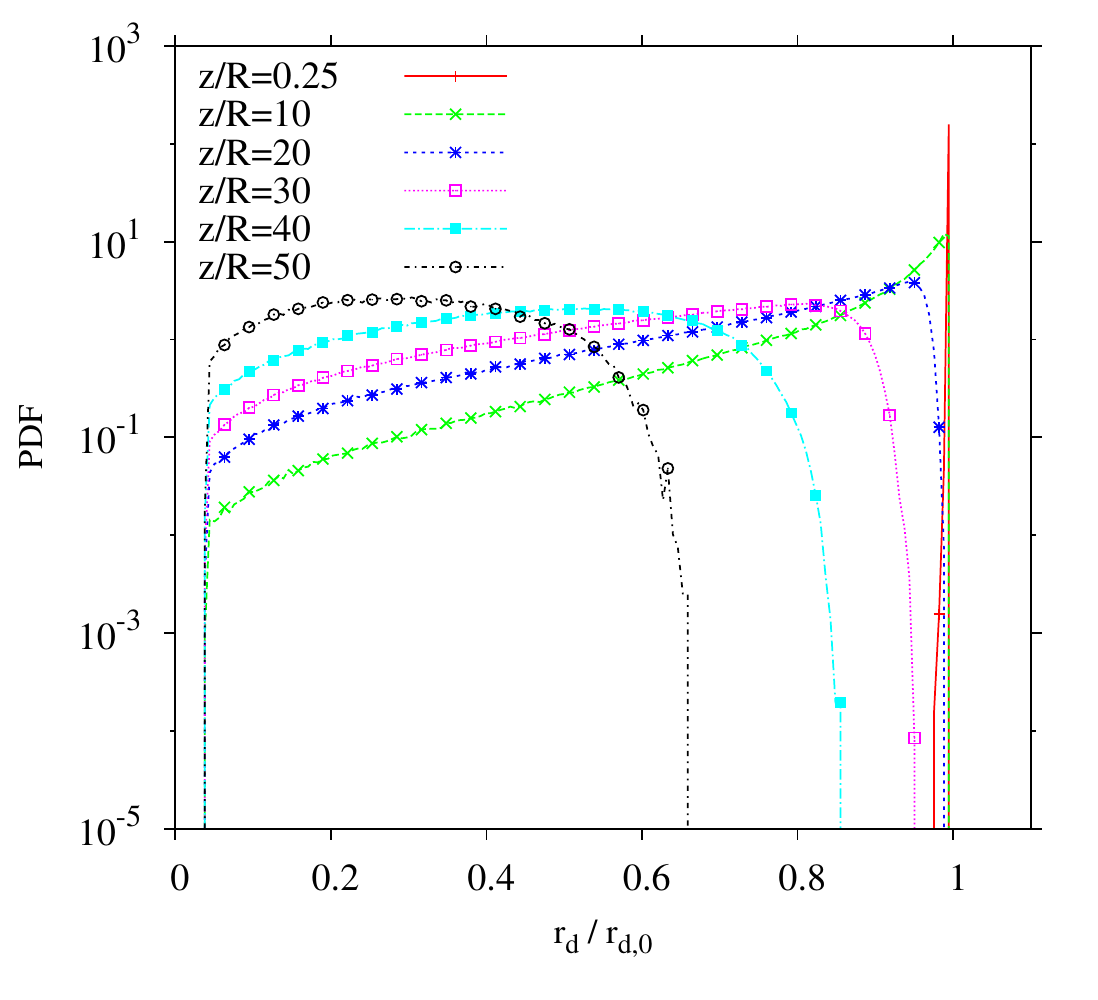} \label{fig:8:a}} \
\subfigure[]{\includegraphics[width=0.45\textwidth]{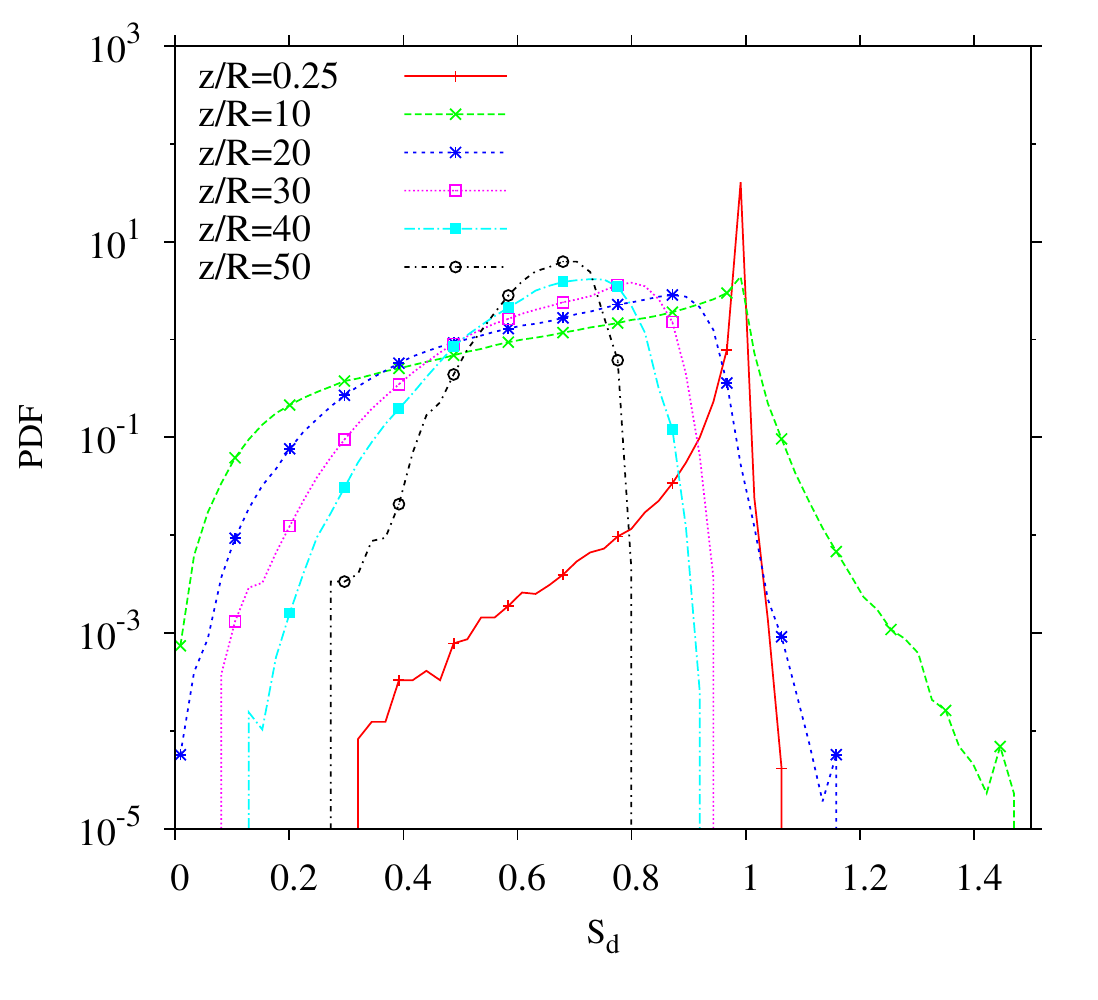} \label{fig:8:b}} \\ 
\subfigure[]{\includegraphics[width=0.45\textwidth]{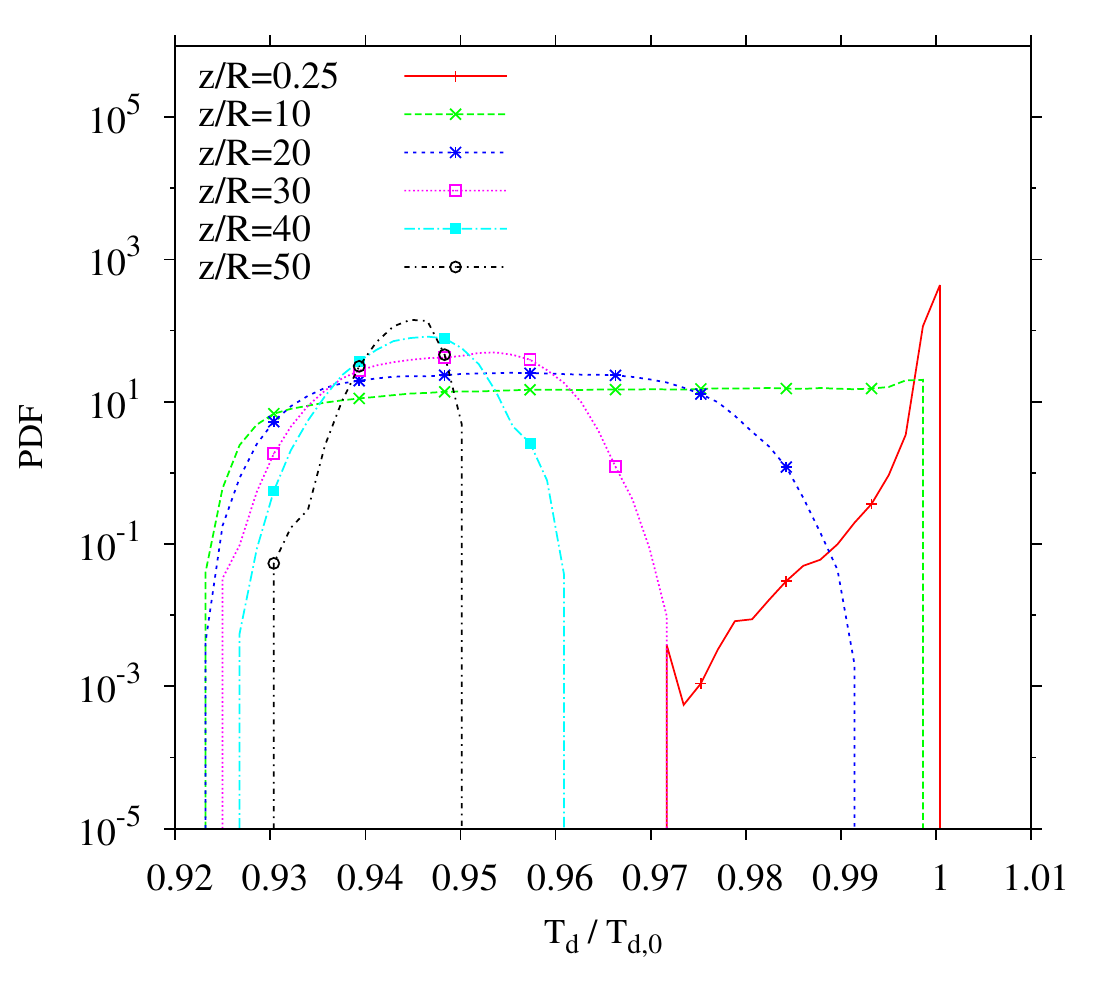} \label{fig:8:c}} \ 
\subfigure[]{\includegraphics[width=0.45\textwidth]{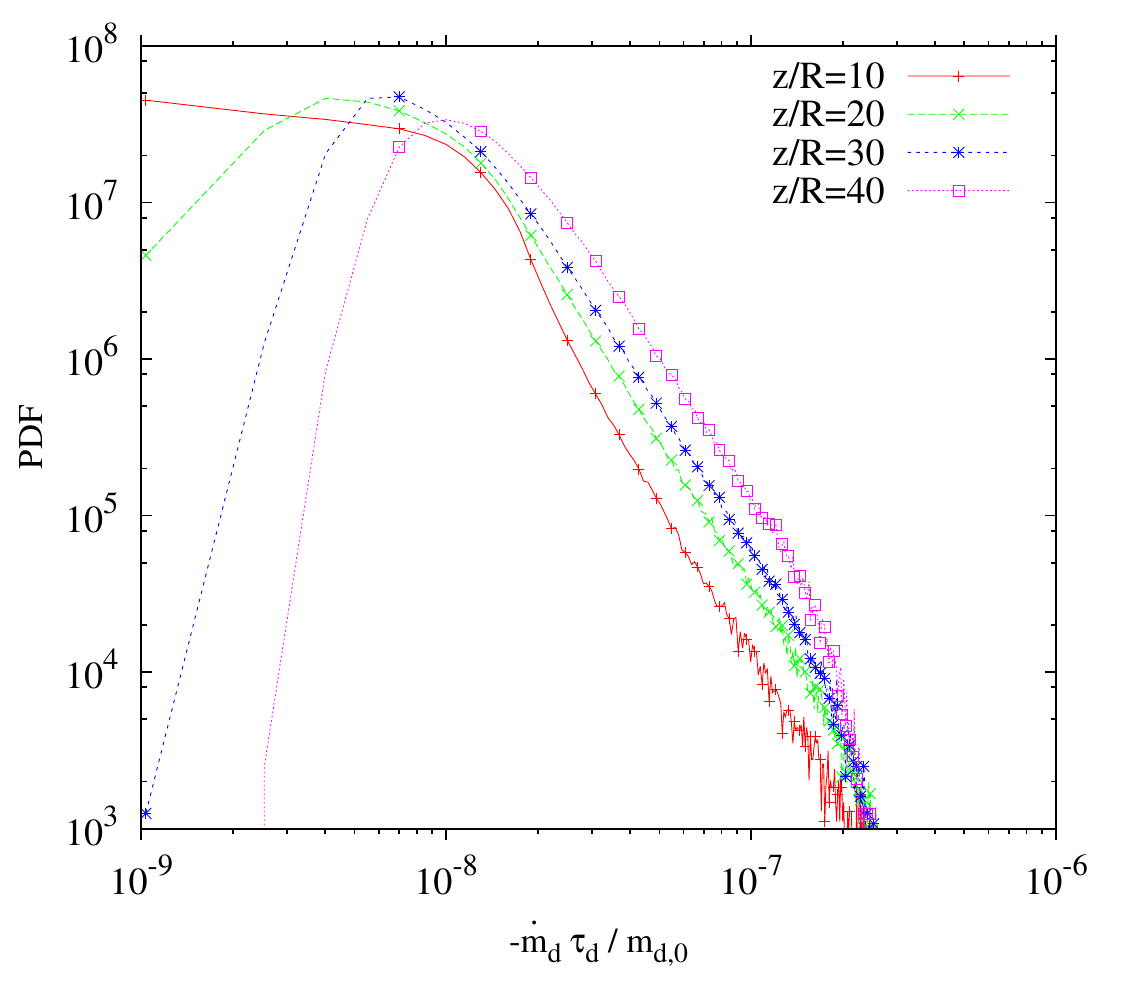} \label{fig:8:d}}
\caption{Probability Density Function (PDF) of Lagrangian variables. (a):  PDF of non-dimensional droplets radius, $r_d/r_{d,0}$, where $r_{d,0}$ is the initial radius of injected droplets. (b): PDF of the saturation field at droplets surface, $S_d=Y_{v,s}/Y_{v,d}$, where $Y_{v,s}=Y_{v,s}(T_d,p)$ is the vapor mass fraction at saturation computed as a function of droplet actual temperature and the carrier phase thermodynamics pressure, p. $Y_{v,ds}$ is the actual vapor mass fraction in the carrier gaseous mixture evaluated at droplet center position. (c): PDF of non-dimensional droplets temperature, $T_d/T_{d,0}$, where $T_{d,0}$ is the initial temperature of injected droplets. (d): PDF of non-dimensional droplets vaporization rate, $-\dot{m_d} \tau_{d,0} / m_{d,0}$, where $\tau_{d,0}$ and $m_{d,0}$ are the initial relaxation time and mass of injected droplets. The PDF plots (a), (b) and (c) are log-linear, while plot (d) is log-log.}
\label{fig:8}
\end{figure*}

Fig. \ref{fig:8:a} shows the probability density function of the droplet radius at different axial distances from the origin. Even starting from a monodisperse suspension, we suddenly observe a radius distribution which spans for around one decade after 10 jet radii from the inlet. It should be remarked that this quantity amounts in differences of droplet volumes of about $10^3$. This intense spread of the droplets size spectrum may be attributed to the complex preferential segregation dynamics which has been previously discussed. Indeed, the evolution of droplets, which is made up of aggregates of different size and dynamics, induces extremely different \emph{surrounding} conditions on droplets themselves as can be observed by the vapor saturation level \emph{felt} by the droplets (see Fig. \ref{fig:8:b}). Near the inlet at $z/R=0.25$, droplets sample the almost saturated vapor phase. From $z/R=10$ the PDF shows a wide range of sampled value caused by the complex droplet dynamics previously discussed. We also observe that few droplets show condensation as denoted by saturation values above 1.  Further downstream $z/R=20$ the spreading trend inverts and droplets are subjected to progressively more uniform saturation levels. The different evaporation dynamics caused by the saturation level felt by the particles induces a similar statistical behavior for the temperature PDF, see Fig. \ref{fig:8:c}.  
The high polydispersity combined to the wide spectrum of saturation levels sampled by the droplets induces a non-trivial behavior of the vaporization rate PDF. Even though we cannot provide arguments, we find that the PDFs of the vaporization rate, shown in Fig.~\ref{fig:8:d}, appears to follow a power-law with exponent about $-3$, independently by the axial distance from inlet. 

\section{Final remarks}

The dynamics of a turbulent evaporating spray is investigated by means of a Direct Numerical Simulation. The  simulation reproduces an acetone/air spray evolving in an open environment considering dilute, non-reacting conditions and accounting for the full coupling between the two phases due to mass, momentum and energy exchange. The entrainment of external dry air is also accounted. Liquid acetone monodisperse droplets are continuously injected within the turbulent gaseous phase at a bulk Reynolds number $Re_R=U R /\nu=6000$. A complete description of both instantaneous and average fields of Eulerian and Lagrangian observables is provided. The distribution of droplets is strongly inhomogeneous with clustering apparent. In particular, droplets seem to preferentially persist in high vapor concentration regions thus being affected by a reduction of the vaporization rate. The intensity of the preferential segregation is estimated by the evaluation of the clustering index. Preferential segregation develops downstream the jet inlet first in the mixing layer and then in the turbulent core. In particular, two different mechanisms driving the inhomogeneous droplet distribution are identified: inertial small scale clustering and droplet segregation induced by the turbulent/non-turbulent interface. The former one is the results of the competition between inertia and Stokes drag and is found to be responsible for droplets preferential accumulation mainly in the spray core and in the far field evolution of the flow. The latter one mainly affects the mixing layer and consists in the entrapment of droplets in  turbulent structures with high vapor concentration which are originated in the core  and protract towards the droplet-free dry environment. Simultaneously, droplet-free dry air regions are engulfed in the jet core enhancing the fluctuation of the droplet concentration. Both these mechanisms affect droplets dynamics and result in an oversampling of the vapor concentration experienced by each droplet, hence affecting the overall vaporization length. Probability Density Function of droplet observable have been reported at different axial distances from the inlet. A spectacular increase of the droplet polydispersity is found to arise in the downstream evolution of the spray resulting in an extreme widening of the droplets size spectrum. This intense spread is attributed to the heavy-tail PDF of the droplet vaporization rate which is the result of the complex dynamics coupling droplet and vapor concentration fields  
This mechanisms is expected to be important in all turbulent flows characterized by a mixing layer with entrainment of dry air, e.g. clouds.\\
The proper modeling of this phenomenon is critical in order to improve LES and RANS model capabilities to accurately reproduce the turbulent vaporization dynamics both for reacting and non-reacting sprays.
\section*{Acknowledgement}

The authors acknowledge financial support through the University of Padova Grant PRAT2015 (CPDA154914), as well as the computer resources provided 
by CINECA ISCRA C project: TaStE (HP10CCB69W).

\bibliographystyle{plain}
\bibliography{biblio}

\end{document}